\input lanlmac
\def\href#1#2{{#2}}

\input epsf.tex

\overfullrule=0mm

\newcount\figno
\figno=0
\def\fig#1#2#3{
\par\begingroup\parindent=0pt\leftskip=1cm\rightskip=1cm\parindent=0pt
\global\advance\figno by 1
\midinsert
\epsfxsize=#3
\centerline{\epsfbox{#2}}
\vskip 12pt
{\bf Fig.\ \the\figno:} #1\par
\endinsert\endgroup\par
}
\def\figtex#1#2{
\par\begingroup\parindent=0pt\leftskip=1cm\rightskip=1cm\parindent=0pt
\global\advance\figno by 1
\midinsert
\centerline{\input #2}
\vskip 12pt
{\bf Fig.\ \the\figno:} #1\par
\endinsert\endgroup\par
}
\def\figlabel#1{\xdef#1{\the\figno}}
\def\encadremath#1{\vbox{\hrule\hbox{\vrule\kern8pt\vbox{\kern8pt
\hbox{$\displaystyle #1$}\kern8pt}
\kern8pt\vrule}\hrule}}


\def\IR{\relax{\rm I\kern-.18em R}}
\font\cmss=cmss10 \font\cmsss=cmss10 at 7pt

\font\cmss=cmss10 \font\cmsss=cmss10 at 7pt
\def\IZ{\relax\ifmmode\mathchoice
{\hbox{\cmss Z\kern-.4em Z}}{\hbox{\cmss Z\kern-.4em Z}}
{\lower.9pt\hbox{\cmsss Z\kern-.4em Z}}
{\lower1.2pt\hbox{\cmsss Z\kern-.4em Z}}\else{\cmss Z\kern-.4em Z}\fi}
\def\IN{\relax{\rm I\kern-.18em N}}
\def\b{\circ}
\def\n{\bullet}

\def\gbbbb{\Gamma_4^{\hbox{$\scriptstyle \b \b$}\kern -8.2pt
\raise -4pt \hbox{$\scriptstyle \b \b$}}}
\def\gnnnn{\Gamma_4^{\hbox{$\scriptstyle \n \n$}\kern -8.2pt  
\raise -4pt \hbox{$\scriptstyle \n \n$}}}
\def\gnnnnnn{\Gamma_6^{\hbox{$\scriptstyle \n \n \n$}\kern -12.3pt
\raise -4pt \hbox{$\scriptstyle \n \n \n$}}}
\def\gbbbbbb{\Gamma_6^{\hbox{$\scriptstyle \b \b \b$}\kern -12.3pt
\raise -4pt \hbox{$\scriptstyle \b \b \b$}}}
\def\gbbbbc{\Gamma_{4\, c}^{\hbox{$\scriptstyle \b \b$}\kern -8.2pt
\raise -4pt \hbox{$\scriptstyle \b \b$}}}
\def\gnnnnc{\Gamma_{4\, c}^{\hbox{$\scriptstyle \n \n$}\kern -8.2pt
\raise -4pt \hbox{$\scriptstyle \n \n$}}}
\def\Rbud#1{{\cal R}_{#1}^{-\kern-1.5pt\blacktriangleright}}
\def\Rleaf#1{{\cal R}_{#1}^{-\kern-1.5pt\vartriangleright}}
\def\Rbudb#1{{\cal R}_{#1}^{\circ\kern-1.5pt-\kern-1.5pt\blacktriangleright}}
\def\Rleafb#1{{\cal R}_{#1}^{\circ\kern-1.5pt-\kern-1.5pt\vartriangleright}}
\def\Rbudn#1{{\cal R}_{#1}^{\bullet\kern-1.5pt-\kern-1.5pt\blacktriangleright}}
\def\Rleafn#1{{\cal R}_{#1}^{\bullet\kern-1.5pt-\kern-1.5pt\vartriangleright}}
\def\Wleaf#1{{\cal W}_{#1}^{-\kern-1.5pt\vartriangleright}}
\def\Cleaf{{\cal C}^{-\kern-1.5pt\vartriangleright}}
\def\Cbud{{\cal C}^{-\kern-1.5pt\blacktriangleright}}
\def\Crleaf{{\cal C}^{-\kern-1.5pt\circledR}}


\magnification=\magstep1
\baselineskip=12pt
\hsize=6.3truein
\vsize=8.7truein
 at 8truept
 at 8truept
 at 10truept

\font\bigrm=cmr12 at 14pt
\centerline{\bigrm Vacancy localization in the square dimer model}

\bigskip\bigskip

\centerline{J. Bouttier$^{1}$, M. Bowick$^{1,2}$,  E. Guitter$^{1}$ and M. 
Jeng$^{2}$}
  \smallskip
  \centerline{${}^{1}$ Service de Physique Th\'eorique, CEA/DSM/SPhT}
  \centerline{Unit\'e de recherche associ\'ee au CNRS}
  \centerline{CEA/Saclay}
  \centerline{91191 Gif sur Yvette Cedex, France}
  \centerline{${}^{2}$ Physics Department}
  \centerline{Syracuse University}
  \centerline{Syracuse, NY 13244-1130, USA}
\centerline{\tt jeremie.bouttier@cea.fr}
\centerline{\tt bowick@phy.syr.edu}
\centerline{\tt emmanuel.guitter@cea.fr}
\centerline{\tt mjeng@phy.syr.edu}

  \bigskip


     \bigskip\bigskip

     \centerline{\bf Abstract}
     \smallskip
     {\narrower
\noindent
We study the classical dimer model on a square lattice with a single
vacancy by developing a graph-theoretic classification of the set of all
configurations which extends the spanning tree formulation of close-packed 
dimers. With this formalism, we can address the question of the possible
motion of the vacancy induced by dimer slidings. We find a probability
$57/4-10\sqrt{2}$ for the vacancy to be strictly jammed in an infinite system.
More generally, the size distribution of the domain 
accessible to the vacancy is characterized by a power law decay with 
exponent $9/8$. On a finite system, the probability that a vacancy 
in the bulk can reach the boundary falls off as a power law of the
system size with exponent $1/4$. The resultant weak localization
of vacancies still allows for unbounded diffusion, characterized by
a diffusion exponent that we relate to that of diffusion on spanning trees. 
We also implement numerical simulations of the model with both free and
periodic boundary conditions.
\par}

     \bigskip


\nref\FORU{R.H. Fowler and G.S. Rushbrooke, {\it Statistical theory
of perfect solutions}, Trans. Faraday Soc. {\bf 33} (1937) 1272-1294.}
\nref\And{P.W. Anderson, {\it The resonating valence bond state in 
La${}_2$CuO${}_4$ and superconductivity}, Science {\bf 235} (1987) 1196-1198.}
\nref\Kast{P.W. Kasteleyn, {\it The statistics of dimers on a lattice. I.
The number of dimer arrangements on a quadratic lattice}, Physica {\bf 27} 
(1961) 1209-1225.}
\nref\KAST{P.W. Kasteleyn, {\it Dimer statistics and phase transitions},
J. Math. Phys. {\bf 4} (1963) 287-293.}
\nref\Fish{M.E. Fisher, {\it Statistical mechanics of dimers on a plane 
lattice}, Phys. Rev. {\bf 124} (1961) 1664-1672.}
\nref\TEFI{H.N.V. Temperley and M.E. Fisher, {\it Dimer problem in 
statistical mechanics - an exact result}, Philos. Mag. (8) {\bf 6} (1961)
1061-1063.}
\nref\FIST{M.E. Fisher and J. Stephenson, {\it Statistical mechanics
of dimers on a plane lattice. II. Dimer correlations and monomers}, 
Phys. Rev. {\bf 132} (1963) 1411-1431.}
\nref\MCWU{B.W. McCoy and T.T. Wu, {\it The two-dimensional Ising model},
Harvard University Press, Cambridge, MA (1973).}
\nref\Hart{R.E. Hartwig, {\it Monomer pair correlations}, J. Math. Phys.
{\bf 7} (1966) 286-299.}
\nref\TSWU{W.J. Tseng and F.Y. Wu, {\it Dimers on a simple-quartic net
with a vacancy}, J. Stat. Phys. {\bf 110} (2003) 671-689.}
\nref\Wu{F.Y. Wu, {\it Pfaffian solution of a dimer-monomer problem: single
monomer on the boundary}, Phys. Rev. E {\bf 74} (2006) 020104(R); Erratum, 
Phys. Rev. E {\bf 74} (2006), 039907(E), arXiv:cond-mat/0607647.}
\nref\Temp{H.N.V. Temperley, in {\it Combinatorics: Proceedings of the British
Combinatorial Conference}, London Math. Soc. Lecture Notes Series No.13 (1974) 
202.}
\nref\KPW{R.W. Kenyon, J.G. Propp and D.B. Wilson, {\it Trees and
matchings}, Elec. J. Comb. {\bf 7} (2000) R25, arXiv:math.CO/9903025.}
\nref\Kirc{G. Kirchhoff, {\it \"Uber die Aufl\"osung der Gleichungen, auf
welche man bei Untersuchung der linearen Verteilung galvanisher Str\"ome
gef\"uhrt wird}, Ann. Phys. Chem. {\bf 72} (1847) 497-508.}
\nref\IPR{N.S. Izmailian, V.B. Priezzhev and P. Ruelle, {\it Non-local
finite-size effects in the dimer model}, SIGMA {\bf 3} (2007) 001 (12 pages),
arXiv:cond-mat/0701075.}
\nref\DUDA{see for instance B. Duplantier and F. David, {\it Exact 
partition functions and 
correlation functions for multiple Hamiltonian walks on the Manhattan 
lattice}, J. Stat. Phys. {\bf 51} (1988) 327-434.}
\nref\Plou{S. Plouffe, http://pi.lacim.uqam.ca/}
\nref\PW{J.G. Propp and D.B. Wilson,
{\it How to get a perfectly random sample from a generic
 Markov chain and generate a random spanning tree of a
 directed graph}, J. Algorithms {\bf 27} (1998) 170-217.}
\nref\PivotAlgorithmOne{W. Krauth and R. Moessner,
{\it Pocket Monte Carlo algorithm for classical doped dimer
 models}, Phys. Rev. B {\bf 67} (2003) 064503.}
\nref\PivotAlgorithmTwo{ W. Krauth,
{\it New Optimization Algorithms in Physics},
eds. A.K. Hartmann and H. Rieger
(Wiley-VCH, 2004), chapter 2.}
\nref\FMS{P. Fendley, R. Moessner and S.L. Sondhi, {\it Classical
dimers on the triangular lattice}, Phys. Rev. B {\bf 66} (2002) 214513.}
\nref\Nien{for an introduction to the Coulomb gas formalism, see
B. Nienhuis, {\it Phase transitions and critical phenomena},
Vol.\ 11, eds.\ C.\ Domb and J.L.\ Lebowitz, Academic Press (1987).}
\nref\Dbw{see also D.B. Wilson's online bibliography at 
http://research.microsoft.com/\~{}dbwilson/exact; first published in 
volume 41 of DIMACS Series in Discrete Mathematics and Theoretical 
Computer Science, published by the American Mathematical Society (1998).}

\newsec{Introduction}

\subsec{The problem}

The statistical mechanics of rigid dimers arranged on a lattice is relevant 
to many physical systems and has appeared repeatedly in the literature over
the years. The study of dimer models can shed light on diatomic gases, the
thermodynamics of adsorbed films \FORU\ and the classical limit of resonating 
valence bond (RVB) models of high-temperature superconductivity \And. 

The case of close-packed dimers on planar lattices is exactly solvable and 
the associated mathematical techniques have proven very powerful 
[\xref\Kast-\xref\MCWU]. It provides
a paradigm of a geometrically constrained statistical system with a strong
interplay between the physical degrees of freedom and the symmetry of the
underlying lattice.
\fig{An example of a dimer move on the square lattice. The dimer 
slides into the empty site (vacancy) of configuration (a), resulting
in a new configuration (b) where the vacancy has jumped by two lattice
spacings.}{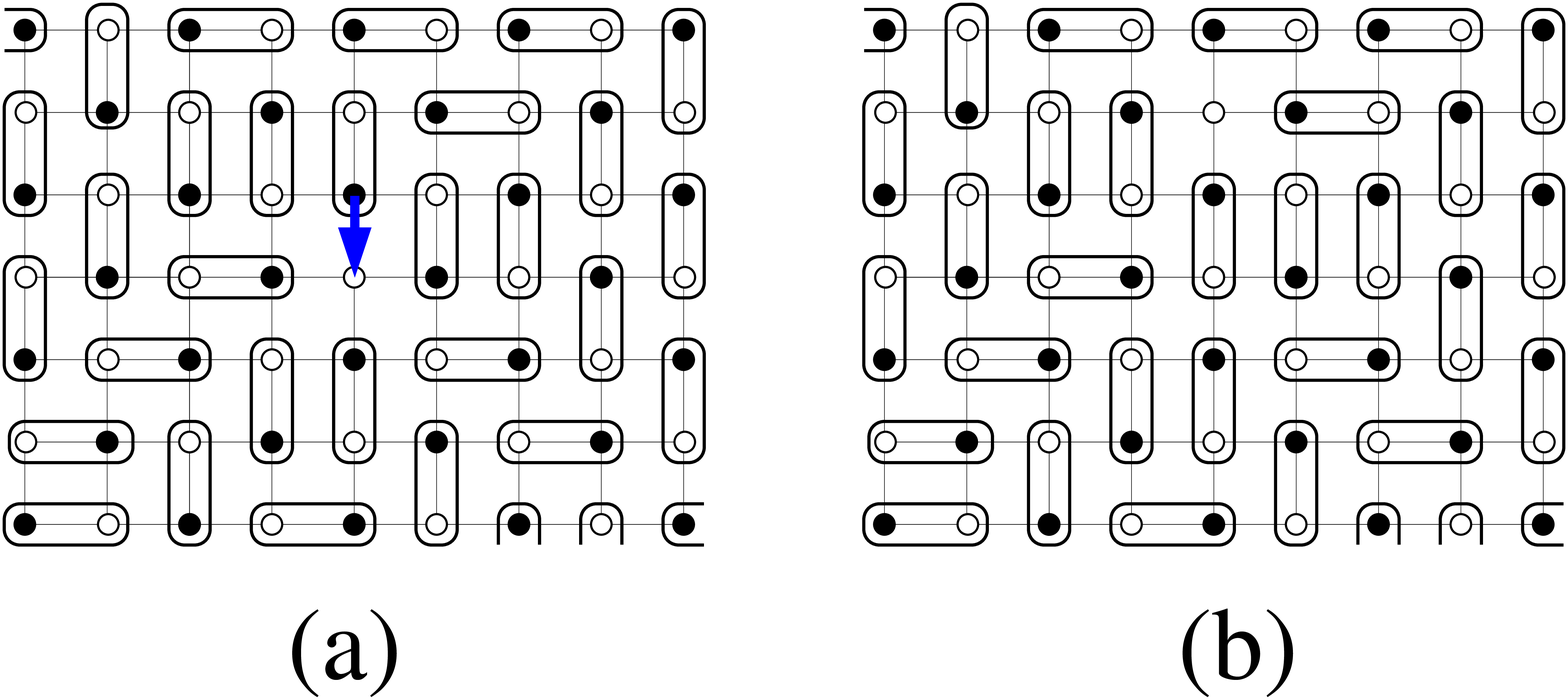}{12.cm}
\figlabel\sliding
New challenges appear if we extend the close-packed dimer models by allowing 
for defects in the form of {\it vacancies}, namely sites not covered by dimers
[\xref\FIST,\xref\Hart-\xref\Wu]. 
The presence of defects allows dynamical moves for dimers by sliding them into 
empty sites, as illustrated in Fig.~\sliding. This in turn induces 
motion of the vacancies in the form of discrete jumps, each by two lattice 
spacings.  It is of great 
interest to characterize as explicitly as possible the nature of the constrained
dynamics of vacancies as a model glassy system. In particular, we would like to
determine the extent to which a vacancy can diffuse, both spatially and 
temporally. 

In this paper, we consider the simplest case of an isolated vacancy in a sea of
dimers on the square lattice. Our primary interest is the structure of the 
space accessible to the vacancy. In particular, we address the question of  
whether or not the vacancy is localized, namely confined to a finite region. 
We find that our model exhibits localization but in a very weak form. 
Although the motion of the vacancy in a fixed dimer background is localized to 
a finite domain, the mean size of this domain nevertheless diverges upon 
averaging over all possible dimer backgrounds. We call this property 
{\it weak localization}, as it allows vacancies to diffuse arbitrarily far on 
average. It is also important to investigate the kinetics of vacancy diffusion. 
We find an anomalous diffusion exponent determined both by the internal 
structure of the space available to the vacancy and the effects of weak 
localization which control the size of this space.

\subsec{Outline and summary of results}

Our paper is structured as follows. In Sect.~2, we recast our dimer problem 
with a single vacancy as a model of spanning graphs on a rectangular grid by 
extending a famous construction by Temperley. This construction is recalled 
in Sect.~2.1 and provides a bijection between dimer configurations with the 
vacancy {\it on the boundary} of the grid and spanning trees on a subgrid. 
The generalization of this construction, in Sect.~2.2, to a vacancy in the bulk 
leads to more general {\it spanning webs}, consisting of a central tree 
component on which the vacancy can freely diffuse, surrounded by a number of 
nested loops with branches, which act as cages for the vacancy. We then 
derive in Sect.~2.3 a determinant formula for the partition function of 
these spanning webs.

Sect.~3 is devoted to a finite size study of dimers on a square grid of 
size $L$, with a vacancy at the center, and the asymptotic limit $L\to \infty$.
In Sect.~3.1, we use our determinant formula to compute the 
probability that the vacancy 
can reach the boundary of the grid, in which case it can reach all sites
of the grid. This ``delocalization probability" is found to decay with the
system size as $L^{-1/4}$. In Sect.~3.2, we analyze the distribution $p(s)$ 
for the size $s$ of the domain accessible to the vacancy in an infinite system. 
In the spanning web formulation, this domain is nothing but the central tree 
component that contains the vacancy. We characterize $p(s)$ by first giving 
the exact value of $p(1)$, which measures the probability that the vacancy 
is strictly jammed. We then derive its large $s$ behavior $p(s) \sim s^{-9/8}$
from the associated scaling of the delocalization probability. 
This power law behavior 
with an exponent larger than $-2$ is responsible for the announced weak 
localization property. We finally discuss in Sect.~3.3 the diffusion exponent 
on spanning webs ($\eta$) and on spanning trees ($\eta_0$) and
find the relation $\eta/\eta_0=7/8$, which measures the slowing 
of diffusion by weak localization.

We complete our analysis in Sect.~4 by a number of numerical simulations both 
on the statistics of spanning webs and the dynamics of diffusion. In Sect.~4.1, 
we show how to modify the well known Propp-Wilson algorithm for the generation
of random spanning trees so as to obtain spanning webs with a uniform measure. 
We present results
for the delocalization probability and for the size distribution $p(s)$,
finding good agreement with the analytic predictions. We check the 
universality of $p(s)$ by also carrying out simulations with periodic boundary 
conditions. There we used the efficient ``pivot algorithm" to generate 
configurations directly in the dimer formulation. 
Sect.~4.2 is devoted to the numerical computation of the diffusion exponents
$\eta_0$ on spanning trees and $\eta$ on spanning webs, whose ratio
agrees with the analytic  prediction.

Appendix A gives a heuristic derivation of the decay exponent 
for the delocalization probability via Coulomb gas arguments.
Appendix B discusses the technical details of the extension of 
the Propp-Wilson algorithm to the generation of spanning webs. 

\newsec{Vacancy in a dense sea of dimers: graph-theoretic treatment}

\subsec{Tree formulation}

\fig{A $9\times 7$ grid (a) is bicolored with one more white than black vertex.
The set of white vertices is made of two intercalated grids of double mesh
size (b) with respective sizes $5\times 4$ and $4\times 3$. The first one
(c) will be referred to as the odd white grid and the second one (d) 
as the even white grid. The latter is extended (dashed lines)
so as to include an extra vertex (here represented by a double boundary)
dual to the exterior face.}{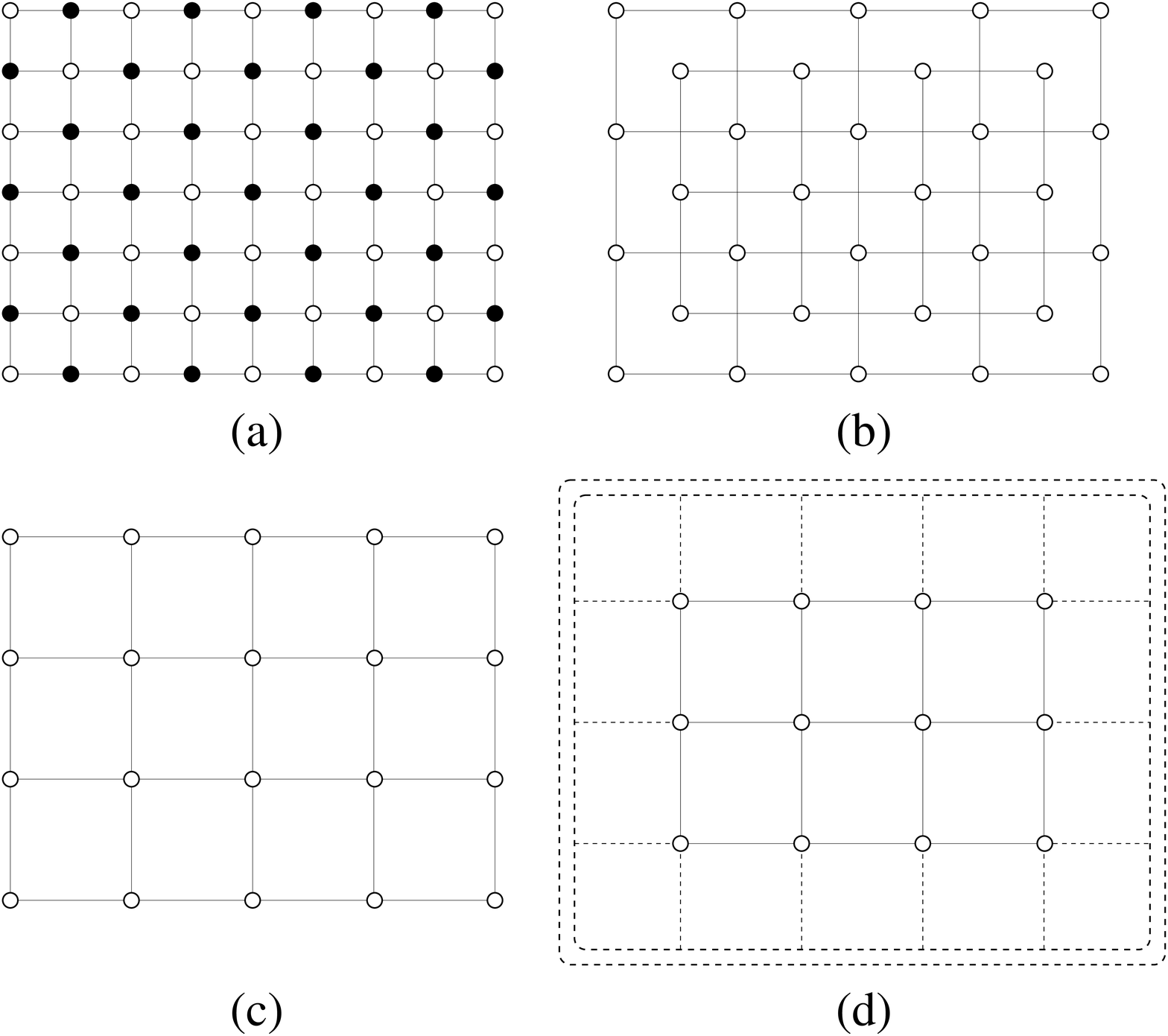}{12.cm}
\figlabel\grid
Let us first recall Temperley's bijection between dimer configurations on a 
rectangular grid and spanning trees on a grid with double mesh size
[\xref\Temp,\xref\KPW]. 
More precisely, let us consider a rectangular grid with $2L+1$ columns
and $2M+1$ rows, so that the total number of vertices is {\it odd}. 
The vertices of the grid can be colored, say in black and white, so that 
neighboring vertices have different colors, with the vertex in the upper right 
corner being white. With this coloring, there is one more white vertex than 
black vertex and the grid can be fully covered by a set of dimers with 
a {\it single} vacancy on a white vertex. Note that the set of white 
vertices can be viewed as made of {\it two intercalated grids}, of size 
$(L+1)\times (M+1)$ and $L\times M$ respectively, both with double mesh size 
(see Fig.~\grid).  We shall refer to these grids as the {\it odd white grid}
and the {\it even white grid} respectively. It is convenient to extend the 
even white grid by an additional white vertex dual to the exterior face of 
the original grid, together with edges from that new vertex to all vertices 
on the boundary of the even white grid (see Fig.~\grid). 
With this convention, 
the ``extended" even white grid is simply the dual graph of the odd white 
grid. Note finally that the black vertices sit precisely on the edges of 
either of these two dual white grids.

\fig{A sample dimer configuration on a $17\times 13$ lattice (a) with a 
single vacancy 
($\bigotimes$) in the upper right corner. In (b) we keep only those dimers
that cover a vertex of the odd white grid and translate them in (c) into
oriented edges of that grid, resulting in a spanning tree configuration 
whose edges are oriented toward the upper right corner. In (d), we also
show the dual tree (red) which spans the extended even white grid and is
oriented to the exterior face. The configuration (a) is recovered by 
replacing every oriented edge of (d) by a dimer on its first 
half.}{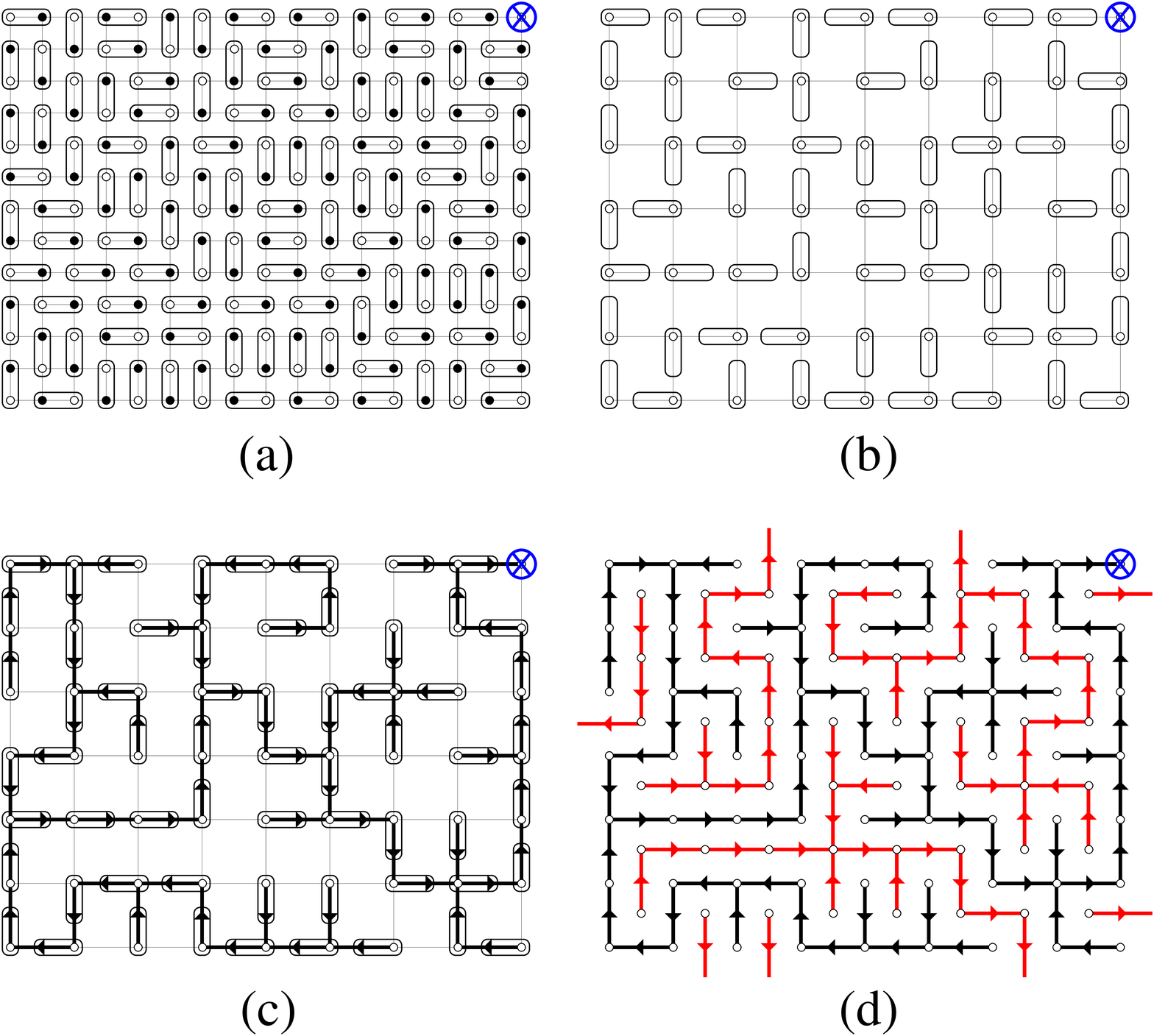}{13.cm}
\figlabel\spanning
Let us assume that we place the vacancy {\it on a white vertex of the boundary}
of the grid, necessarily part of the odd white grid. Now any other vertex of 
the odd white grid carries a dimer (see Fig.~\spanning-(a)). 
This dimer selects an edge of the odd white
grid, which we decide to mark and orient from the white vertex carrying that 
dimer to its neighbor on the odd white grid (see Fig.~\spanning-(b,c)). 
The graph made of all these marked
oriented edges together with all the original vertices of the odd white grid
forms a spanning tree of the odd white grid whose edges are moreover oriented 
toward the vacancy (see Fig.~\spanning-(d)). 
To understand the absence of loops in the graph, note 
that a loop would enclose an interior region of the original lattice 
with an odd number 
of vertices that therefore could not be fully covered by dimers (here it is 
crucial that we have taken the vacancy to lie strictly on the boundary). 
As the graph 
has exactly one more vertex than edges (as the vertex carrying the vacancy does 
not give rise to a marked edge), it is necessarily made of a single tree
spanning the whole odd white grid. 
The edge orientations produce the unique flow on the tree toward the vacancy 
(with exactly one edge exiting from all vertices but the vacancy). 
In particular, if we fix the 
position of the vacancy, say in the upper right corner, the above construction 
provides a {\it bijection} between fully-packed dimer configurations of a 
$(2L+1) \times (2M+1)$ grid with one vacancy in the corner and spanning trees 
of the associated $(L+1)\times (M+1)$ odd white grid.
To go from the spanning tree configuration back to the original 
dimer configuration, we first orient each edge of the tree so as to reproduce 
the unique flow toward the upper right corner and replace each edge of the 
tree by a dimer on the first half of the edge. We then consider 
the ``dual tree" of the spanning tree, made of those edges of the extended 
even white grid that do not cross the edges of the spanning tree. 
This dual graph is itself a tree spanning the extended even white grid
which we orient toward the exterior vertex (see Fig.~\spanning-(d)). 
We finally repeat the above 
construction and replace each oriented edge of the dual tree by a dimer 
on the first half of the edge. 

Clearly, we have a similar bijection if we place the vacancy at some arbitrary 
but {\it fixed} white vertex on the boundary of the grid. This simply amounts 
to consider spanning trees with another orientation of the edges, now pointing 
toward the new position of the vacancy on the boundary.
If we leave the position of the vacancy free along the boundary, we 
clearly get a $2(L+M)$ to $1$ mapping instead. 

\fig{The elementary motion of the vacancy resulting from the sliding of a 
dimer (top). In the tree formulation, this amounts to reversing the orientation
of the corresponding edge on the (oriented) spanning tree.}{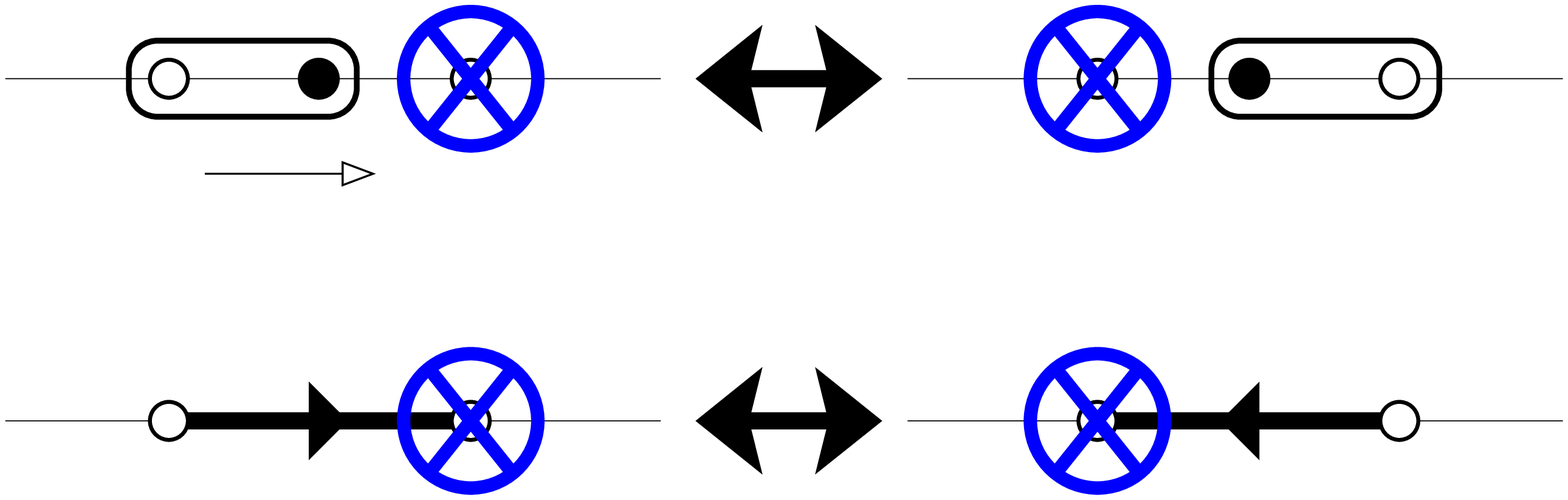}{7.cm}
\figlabel\move
Let us now consider the motion of the vacancy generated by sliding of dimers. 
In the tree formulation, as illustrated in Fig.~\move, performing an 
elementary slide corresponds to picking 
an edge adjacent to the vacancy (and necessarily pointing to it) and reversing 
its orientation so that the flow now points to a new vertex. We can then repeat
the process, which allows the vacancy to reach any vertex on the
spanning tree, hence {\it any vertex} of the odd
white grid. In the tree formulation, the dimer configurations accessible by 
slidings are therefore described by the {\it same spanning tree} and differ 
only by the choice of the (arbitrary) vacancy site on the odd white grid 
toward which we orient all the edges of this tree. Each orientation 
may then be mapped into dimers as above. Note that the dimers corresponding 
to edges of the dual tree cannot be affected by slidings and are in practice 
frozen. Note finally that the
property that the vacancy can reach any site of the odd white grid holds
only because the vacancy was on the boundary in the first place. As
shall see just below, this property is not true in general for vacancies 
that lie inside the grid. To conclude, the motion of the vacancy 
can be analyzed as simple diffusion on a spanning tree picked uniformly
at random.

\fig{A sample dimer configuration (a) with a single vacancy 
($\bigotimes$) in the bulk of the odd white grid. In (b) we show only 
those dimers that cover a vertex of the odd white grid. When mapping
them into oriented edges of that grid, the resulting spanning web (c) 
consists of a tree component containing the vacancy, surrounded by a number 
of loop components (here $2$). The edges belonging to branches of the tree 
component are oriented toward the vacancy while those belonging to branches of 
the loop components are oriented toward the loop. On each loop, all edges 
have the same orientation. In (d), we also show the ``dual web" (red) which 
spans 
the extended even white grid and is made of an equal number of loop components
surrounding the vacancy together with an extra tree component attached
to the exterior face. The edges belonging to branches of this tree component 
are oriented toward the exterior face while those belonging to branches 
of the loop components are oriented toward the loop. Again, all edges have the 
same orientation on each loop. The configuration (a) is recovered by 
replacing every oriented edge of (d) by a dimer on its first 
half.}{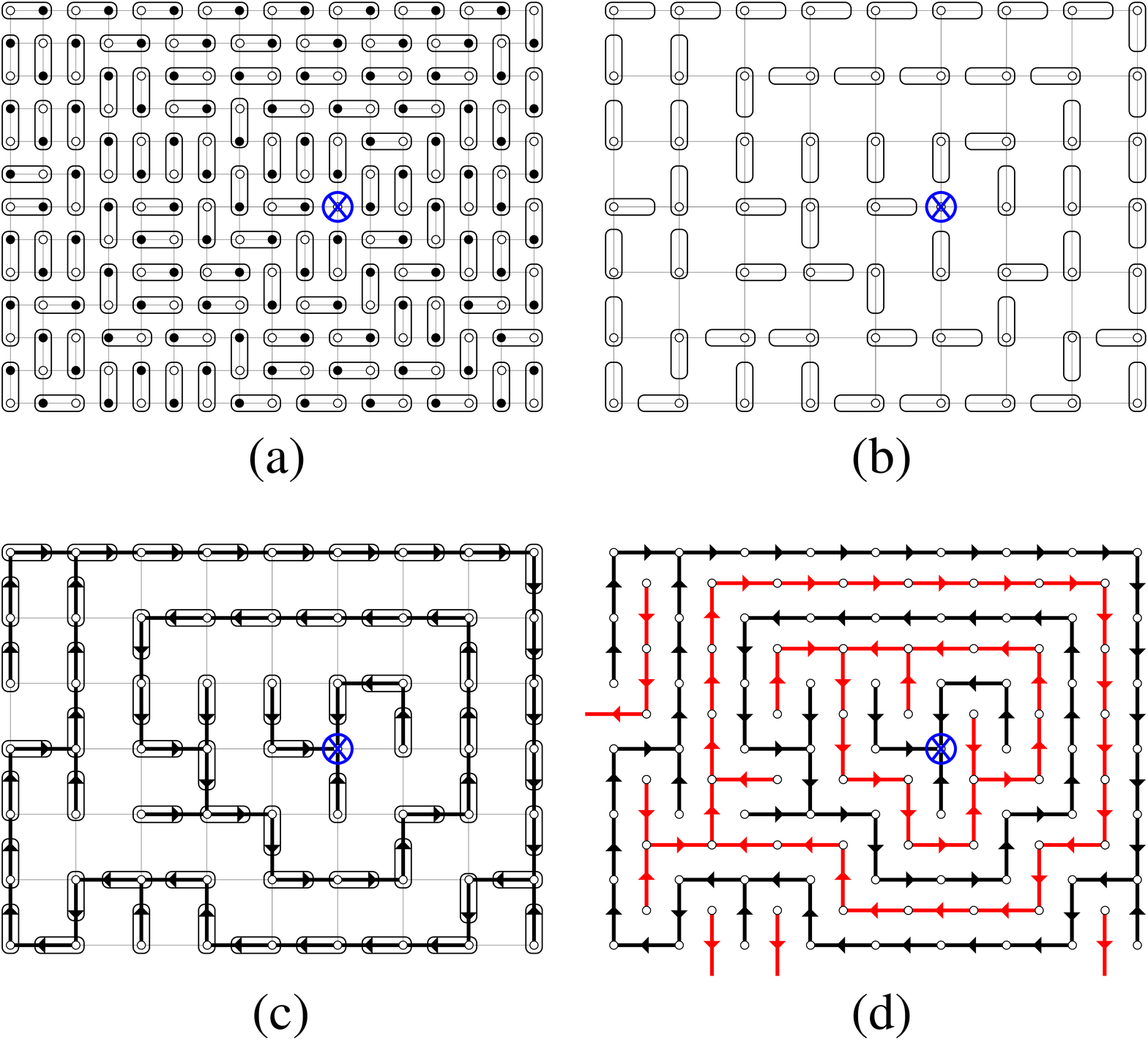}{13.cm}
\figlabel\web

\subsec{Extension to webs}

Let us now consider the more general case where the vacancy originally lies 
{\it on 
an arbitrary vertex of the odd white grid}, not necessarily on the boundary
(see Fig.~\web-(a)). 
We can repeat the above construction by marking and orienting, for each vertex 
of the odd white grid except the vacancy vertex, the edge that 
is selected by the 
dimer covering that vertex (see Fig.~\web-(b)). As before, we consider 
the graph made of all these
marked oriented edges together with all the original vertices of the odd white
grid. As there is exactly one oriented edge exiting from each vertex but the
vacancy vertex, any connected component of this graph is either a tree 
containing the vacancy vertex and with edges pointing to that vertex or it 
is made of an oriented loop with attached branches oriented toward it 
(see Fig.~\web-(c)). In
this latter case, the connected component cannot contain the vacancy vertex.
Moreover, as before, any such loop encloses an odd number of vertices, so {\it 
it must encircle the vacancy}, which is possible only if the vacancy does
not lie on the boundary. To summarize, the general structure of the graph
is a tree containing the vacancy vertex (the tree may be as small as a single 
vertex) surrounded by a number of nested loop components (made of a loop
and attached branches) so that the graph
spans the entire odd white grid. We shall call such a structure a 
``spanning web" of the odd white grid rooted at the vacancy vertex. 

Conversely, given such a graph, we note that the 
``dual web" made of those edges of the extended even white grid that do not 
cross the edges of the spanning web is itself a spanning web of the
extended even white grid (see Fig.~\web-(d)). It is made of a set of 
nested loop components, 
with branches, complemented by a tree rooted at the exterior vertex. The loops
of the dual web and that of the original spanning web alternate and therefore
match in number. We can translate the spanning web configuration into
a dimer configuration by successively orienting the edges of its tree 
component toward one of its vertices, where we put the vacancy, the edges
of each loop in the same direction, chosen arbitrarily for each loop, 
and finally the edges of the attached branches toward the loops. Similarly, 
we orient the edges of the tree component of the dual web toward the exterior 
vertex, those of the loops in the same direction, arbitrarily for each loop 
and those of the attached branches toward the loops. The dimer configuration
is obtained by replacing each of these oriented edges by a dimer on 
the first half of the edge.

The above construction provides a bijection between dimer configurations
with a vacancy at a fixed position on the odd white grid and spanning
webs whose tree component contains the vacancy, together with a choice of
orientation for each loop of the spanning web and each loop of the dual web.
As there are 2 possible orientations per loop and an equal number of loops 
on both web configurations, we can get rid of the orientations by counting
each loop of the original spanning web with a degeneracy factor 4.

Again, let us examine how the vacancy can move under the sliding of dimers. As
before, an elementary slide amounts in the spanning web language to reversing 
the orientation of one edge pointing to the vacancy so that it points to a
new vertex. Under repeated moves, the vacancy can reach every white site of 
the tree component of the spanning web, which therefore constitutes a 
complete specification of the set of sites accessible to the vacancy.
Note that all the dimers corresponding to oriented edges of either the loop 
components of the spanning web or any component of its dual web are
frozen. In particular, the vacancy cannot cross a loop of either web.

On a finite grid, we may therefore divide the configurations into two classes, 
depending on whether or
not the vacancy can reach the boundary. If the vacancy can reach the boundary,
there cannot be any loop as the loops are required to encircle the tree 
component of the spanning web. In this case, the spanning web reduces to
a spanning tree and the vacancy can reach any vertex of the odd white grid.
If the vacancy cannot reach the boundary, there must be a loop component
containing all the white boundary vertices. There are in general several
nested loops, the interior-most one acting as a cage for the vacancy. 
The motion of the vacancy may then be analyzed as simple diffusion on
the tree component of the web.
For a {\it fixed initial position} of the vacancy on the grid, the ratio 
between the number of dimer configurations for which the vacancy can reach 
the boundary and the total number of dimer configurations can be interpreted 
as a delocalization probability. This probability will be studied in detail
in Sect.~3.1 below for the case of a square grid with vacancy in the 
center. 

On an infinite grid, the vacancy is always localized to a finite tree
component but we shall see that the mean size of this tree in fact diverges.

\subsec{Determinant formulae}

It is well known that the number of spanning trees on a graph with $n$ vertices 
is given by any principal $(n-1)\times (n-1)$ minor of the Laplacian matrix
of this graph. This is the celebrated matrix-tree theorem attributed to 
Kirchhoff \Kirc. Recall that 
the coefficients of the Laplacian matrix simply read
\eqn\laplacian{\Delta_{ij}= \left\{\matrix{\hfill d_i & \ \hbox{for $i=j$}
\hfill\cr \hfill -1 & \ \hbox{for $i$ and $j$ neighbors} \hfill 
\cr \hfill 0 & \ \hbox{otherwise \ , } \hfill \cr}\right.}
where $i$ and $j$ are vertices of the graph and $d_i$ is the degree of $i$ 
(number of incident edges) on the graph.

Here we are interested in the case of a rectangular grid where $d_i$ can be
$2$, $3$ or $4$ according to whether $i$ lies at a corner, on the side
or in the bulk of the grid. Computing the minor amounts to removing a given
vertex $i_0$, i.e. restricting the indices of the matrix to all $i\neq i_0$.
Note that although $i_0$ is removed, the degrees of its neighbors are kept
unchanged. The number $Z_{\rm tree}$ of spanning trees therefore reads
\eqn\determinant{Z_{\rm tree}=\det \left(\Delta_{ij}\right)_{i,j\neq i_0}\ .}
\fig{A graphical representation of a term in the expansion of the determinant
in Eq.~\determinant\ by 
use of oriented blue and red edges. There is exactly one 
outgoing arrow from every vertex distinct from $i_0$. Blue edges must form 
oriented cycles, each cycles receiving a weight $-1$. Red edges may either
form oriented cycles or branches oriented toward a (blue or red) cycle or
toward $i_0$.}{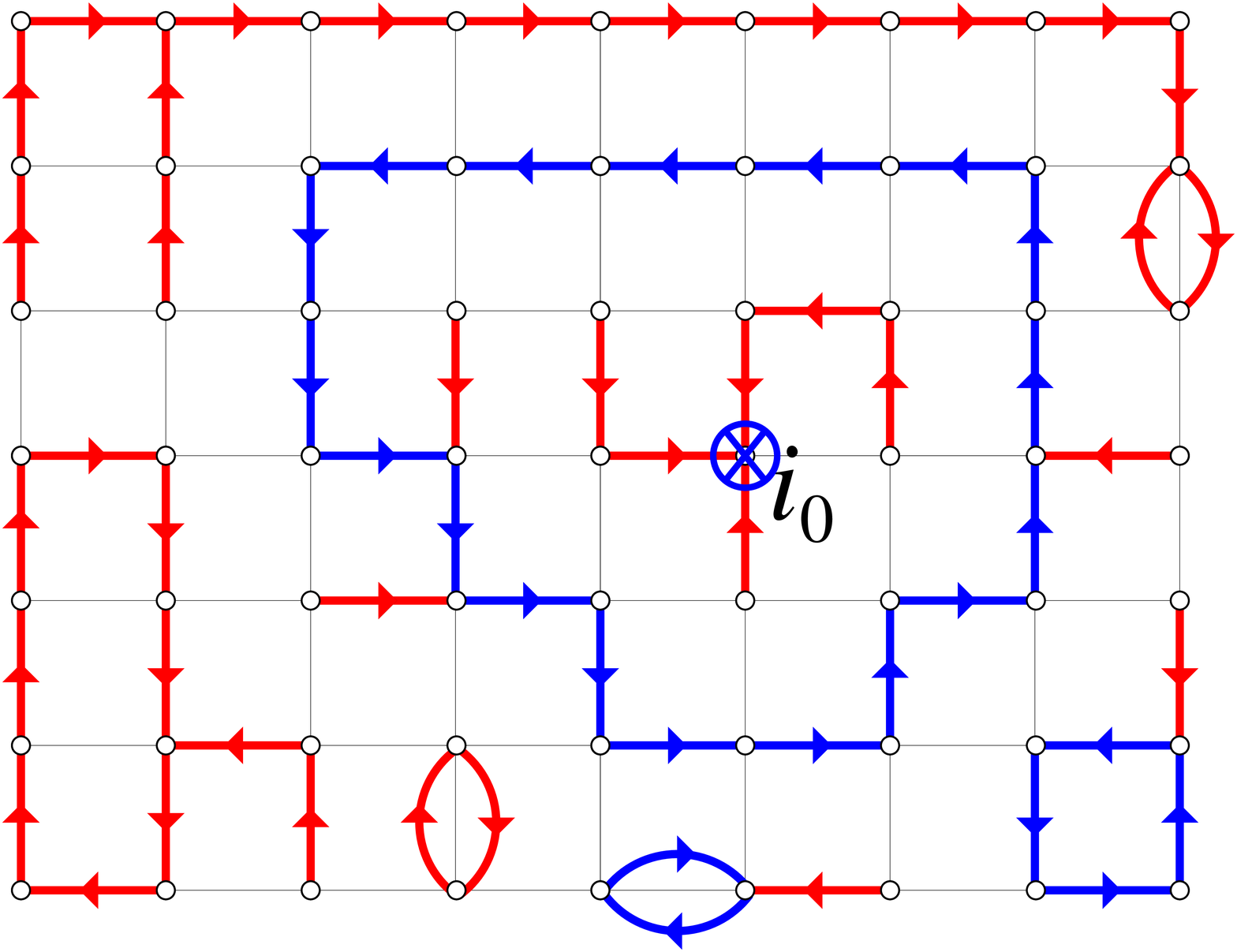}{7.cm}
\figlabel\detfig

Let us sketch a proof of this result, along the lines of Ref.~\IPR,
that can be easily extended to spanning webs. Writing the determinant 
as a sum over
permutations, we see that the only permutations having a non-zero 
contribution are those with the following two types of cycles: (i) trivial 
cycles of length one (fixed points), each contributing a weight 
equal to the degree the associated vertex,  (ii) cycles of length larger 
or equal to $2$ for which any two successive elements of the cycle are 
neighbors on the grid. The net contribution of any such non-trivial cycle 
to the determinant (including the signature of the permutation) is 
easily seen to be $-1$. It is convenient to have a pictorial representation
of each such permutation as follows. Any cycle of type (ii) above
can be represented as a closed oriented loop on the grid by joining
each vertex to its image under the permutation. The loops moreover avoid
the removed vertex $i_0$. 
Such loops are represented in blue in Fig.~\detfig\ and will be henceforth
referred to as blue loops. Each blue loop contributes a factor $-1$. 
For the trivial cycles (i), the contribution $d_i$ of 
the associated vertex $i$ is properly accounted for by considering the
$d_i$ possible choices of an edge incident to that vertex. For each choice, 
we mark and orient the edge away from the vertex. Such oriented edges are
represented in red in Fig.~\detfig\ and will be referred to as red edges.
Any permutation is thus represented by a number of configurations 
(corresponding to the different choices of red edges for trivial cycles) made
of blue loops and red edges with the only constraint that there is exactly
one outgoing (blue or red) edge from each vertex of the grid but the
removed vertex $i_0$.  Each configuration is now weighted by a factor 
$1$ per red edge and $-1$ per blue loop. Now clearly, any configuration in
which the red edges form a loop is canceled exactly by a similar
configuration where all the edges of this loop are blue. The only 
configurations that remain therefore consist of red edges forming
a spanning tree, necessarily oriented toward the removed vertex $i_0$.
This completes the proof.

From the determinant formula in Eq.~\determinant, one can easily derive 
a closed formula for $Z_{\rm tree}$ as a product over eigenvalues
of $\Delta$ \DUDA.

\fig{The odd white grid (a) completed with a seam from $i_0$ to the
boundary (thick dashed line). This seam crosses a number of edges that
we call seam edges and that we orient so as to point to the right when
flowing on the seam from $i_0$ to the boundary. 
In the graphical expansion of the determinant (b), the seam  
modifies the weight of the blue edges crossing it, resulting in a new
weight for blue loops. Here the blue loop that winds
around $i_0$ passes through exactly one seam edge (with the wrong orientation),
resulting in a weight factor $-1/a$ instead of $-1$ in the absence of a seam.
On the contrary, the lower-left blue loop that does not wind around $i_0$ passes
exactly twice through seam edges, with canceling orientations so that 
its weight is unchanged.}{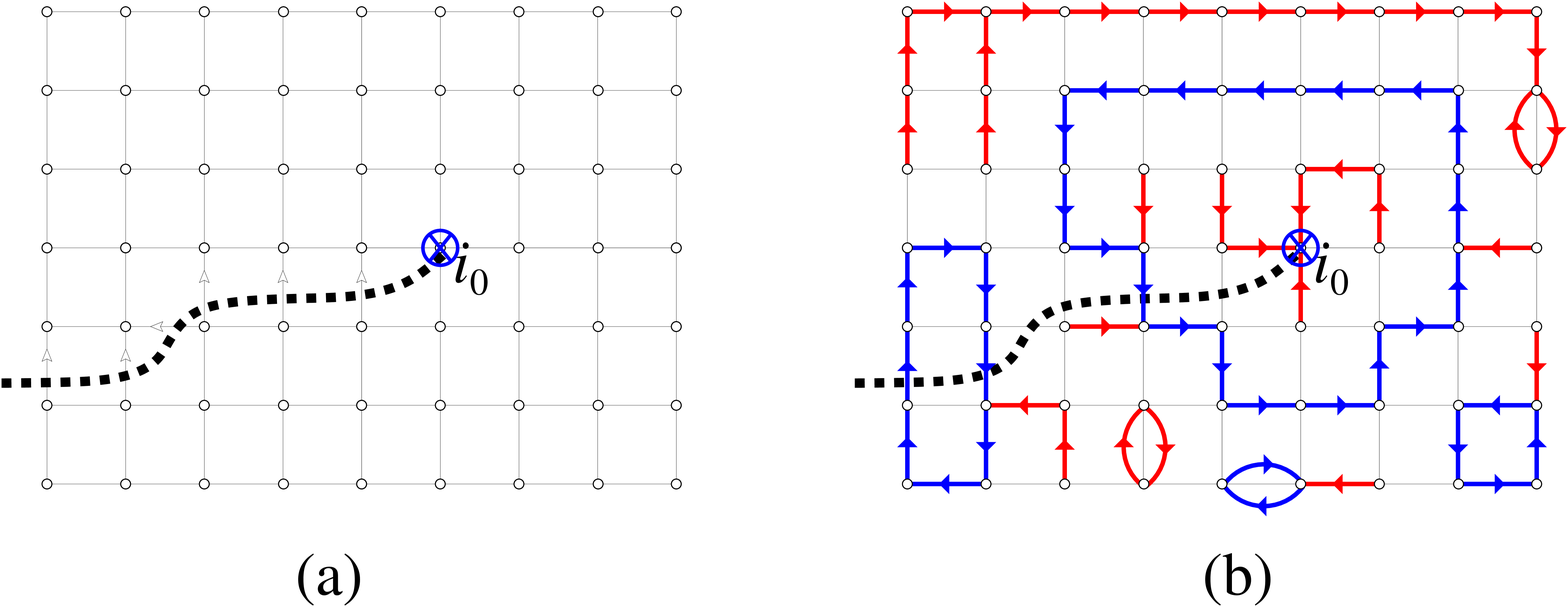}{13.cm}
\figlabel\seam

The expression \determinant\ is easily modified to obtain the number 
$Z_{\rm web}(i_0)$ of spanning webs on a rectangular grid rooted at some
fixed vertex $i_0$. We simply construct an oriented seam from $i_0$ 
to the exterior face that crosses edges only (see Fig.~\seam). Each
crossed edge, which we denote a seam edge, can be oriented so as to point 
right when flowing along the seam.
We may then modify the Laplacian matrix into a matrix $\Delta(a)$ defined
as:
\eqn\modlap{\Delta(a)_{ij}= \left\{\matrix{\hfill d_i & \ \hbox{for $i=j$}
\hfill\cr 
\hfill -1 & \ \hbox{for $i$, $j$ neighbors not separated by
a seam edge} \hfill \cr 
\hfill -a & \ \hbox{for $i$, $j$ neighbors separated by
a seam edge oriented from $i$ to $j$} \hfill \cr 
\hfill -{1\over a} & \ \hbox{for $i$, $j$ neighbors separated by
a seam edge oriented from $j$ to $i$} \hfill \cr 
\hfill 0 & \ \hbox{otherwise \ .} \hfill \cr}\right.}
For any choice of the seam, the quantity
\eqn\moddet{Z(y;i_0)\equiv \det \left(\Delta(a)_{ij}\right)_{i,j\neq i_0}}
then counts the number of spanning webs with root $i_0$ and with 
a weight 
\eqn\atoy{y=2-a-{1\over a}}
per loop. In particular, we recover $Z_{\rm tree}=Z(0;i_0)$ for $a=1$ while
for spanning webs, the desired weight $4$ per loop is obtained by 
choosing $a=-1$, i.e. 
\eqn\zweb{Z_{\rm web}(i_0)=Z(4;i_0)=
\det \left(\Delta(-1)_{ij}\right)_{i,j\neq i_0}\ .}
To obtain this result, we simply use for the determinant the same 
representation as above in terms of blue loops and red edges.
Defining the algebraic number of seam crossings as the number of passages
across seam edges in the correct orientation minus that in the wrong
orientation, this number is zero for a loop that does not encircle $i_0$,
$+1$ for a loop that encircles $i_0$ clockwise, and $-1$ for a loop
that encircles $i_0$ counterclockwise.  The blue loops that wind around 
the vertex $i_0$ now get a modified weight $-a$ or $-1/a$ according to 
whether they are 
oriented clockwise or counterclockwise. As the weight of red edges
is unaffected, any clockwise (resp. counterclockwise) red loop encircling 
the vertex $i_0$ combined with the same configuration with a blue loop 
instead gives rise to a weight $1-a$ (resp. $1-1/a$). Summing over 
both orientation results in the weight $y$ above.

\newsec{Finite size analysis}

\subsec{Determinant calculations and asymptotic estimates}

\fig{The reduction of the odd white grid (a) to one of its quadrants (b)
completed by oriented winding edges.}{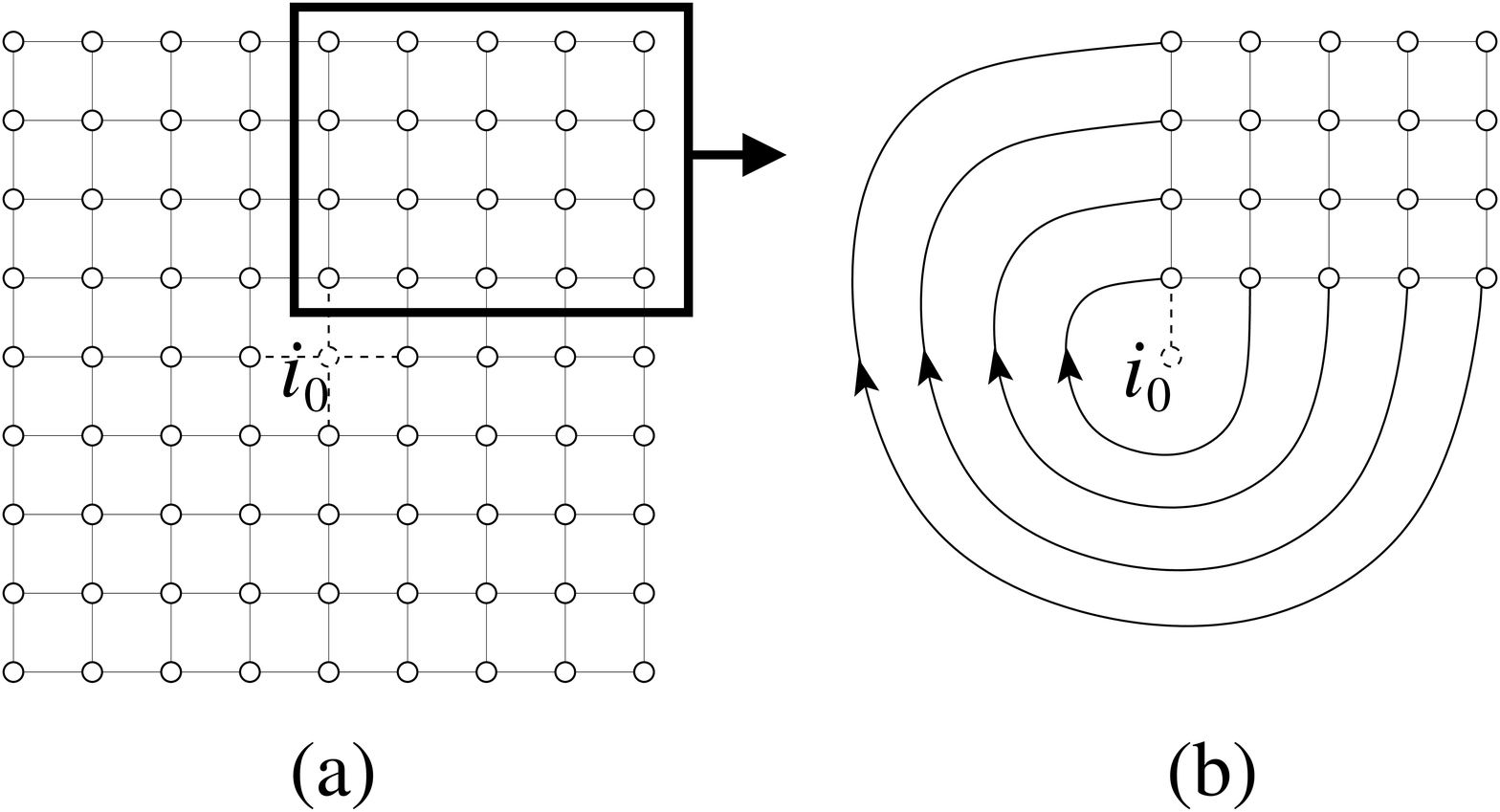}{13.cm}
\figlabel\quarter

In this section, we shall restrict our analysis to the case of dimers on a
square grid with the vacancy in the center. We shall consider only squares
of size $(4L+1)\times (4L+1)$ so that the center vertex $i_0$ lies on the 
odd white grid, of size $(2L+1)\times(2L+1)$. With this geometry, we can rely
on the fourfold rotational symmetry of the problem to perform a 
block-diagonalization of $\Delta(a)$ which translates into the factorization:
\eqn\factorZ{Z(y;i_0)=P_L(\alpha)P_L(-\alpha)P_L(\hbox{i}\,\alpha)
P_L(-\hbox{i}\,\alpha)\ ,}
where $\alpha^4=a$ with $a$ as in Eq.~\atoy, namely 
\eqn\alphatoy{y=2-\alpha^4-{1\over \alpha^4}\ .}
Here $P_L(\alpha)$ is the determinant of a square matrix 
${\bar \Delta}(\alpha)$ of size $L(L+1)$ 
(compared to $4L(L+1)$ for $\Delta(a)$) corresponding to a modified 
Laplacian on a {\it quarter grid}. More precisely, we consider the graph of
Fig.~\quarter\ obtained by keeping only one quadrant of the original odd 
white grid and completing it by adding $L$ oriented ``winding edges" 
joining the vertices of two consecutive sides of the quadrant as shown. 
The matrix 
${\bar \Delta}(\alpha)$ has diagonal elements ${\bar\Delta}(\alpha)_{ii} = 
d_i$ where $d_i$ is the degree of site $i$ on the graph of Fig.~\quarter\
completed by the vertex $i_0$ as shown, which is also the degree
of the corresponding vertex on the original odd white grid. The 
off-diagonal elements ${\bar\Delta}(\alpha)_{ij}$ are given by the sum over 
all edges connecting $i$ and $j$ of a contribution equal to $-1$ for regular
(non-winding) edges, $-\alpha$ (resp. $-1/\alpha$) for winding edges oriented
from $i$ to $j$ (resp. from $j$ to $i$).

It is interesting to note that $P_L(\alpha)$ can be interpreted as the
number of spanning web configurations rooted at the center of the square
grid of size $(2L+1)\times(2L+1)$ and which are {\it symmetric}
under $\pi/2$ rotations around the vacancy vertex.
For any such symmetric spanning web, the loops now come with a weight 
$y=2-\alpha-1/\alpha$. In particular, for $\alpha=-1$, $P_L(-1)$ counts the
number of fourfold symmetric dimer configurations on the square grid
of size $(4L+1)\times(4L+1)$. 

\figtex{The numerical value (a) of $\log_{10}(\pi_{L,i})$ versus $i$ for 
$L=10$, $20$, $30$, $40$, $50$, and $60$, from bottom to top. All the
data fall on the same scaling curve (b) upon using reduced variables
$i/L$ and $\log_{10}(\pi_{L,i})/\log_{10}(\pi_{L,0})$.}
{coeffpl.tex}
\figlabel\coeffpl

Using Mathematica, we have computed $P_L(\alpha)$ exactly up to $L=60$, 
corresponding to dimers on a $241\times 241$ grid. 
For $L=1,2,3$, we have 
\eqn\firstP{\eqalign{
P_1(\alpha) &=4-\left(\alpha+{1\over \alpha}\right) \cr 
P_2(\alpha) &=178-60\left(\alpha+{1\over \alpha}\right)+\left(
\alpha^2+{1\over\alpha^2}\right) \cr 
P_3(\alpha) &=82128-31667\left(\alpha+{1\over \alpha}\right)+1160
\left(\alpha^2+ {1\over\alpha^2}\right) -\left(\alpha^3+{1\over\alpha^3}
\right)\ .\cr}}
More generally, $P_L(\alpha)$ is a Laurent polynomial of $\alpha$ of the form
\eqn\formPL{P_L(\alpha)=\sum_{i=-L}^{L} \pi_{L,i} (-\alpha)^i}
where the $\pi_{L,i}$ are positive integers and satisfy 
$\pi_{L,-i}=\pi_{L,i}$. Their numerical values for $L=10$, $20$, $30$,
$40$, $50$ and $60$ are displayed in Fig.~\coeffpl. 

From our data, we first conjecture the amusing fact 
that $P_L({\rm i})$ counts the number of fully packed dimer 
configurations on a cylinder of height $2L$ and circumference $2L+1$.
 
More relevant to our study, we expect a large $L$ behavior
\eqn\scalP{P_L(\alpha)\sim \tilde c(\alpha) {\mu^{L(L+1)}\, \lambda^L \over 
L^{\tilde\gamma(\alpha)}}\ ,}
where the area entropy factor $\mu$ and the boundary entropy factor $\lambda$
are independent of $\alpha$ while the prefactor $c$ and the exponent
$\tilde\gamma$ depend on $\alpha$. The values of $\mu$ and $\lambda$ are known
to be \DUDA 
\eqn\valentropy{\eqalign{\mu &= \exp \left({1\over 4\pi^2} \int_0^{2\pi}
dk \int_0^{2\pi} d\ell \ \ln\left(4-2\cos k-
2\cos \ell\right)\right)\cr &= \exp \left({4 G \over \pi}\right) 
=3.209912300728158\dots \cr
\lambda &= \sqrt{2}-1 \ ,\cr}}
where $G=\sum_{i=0}^\infty (-1)^i/(2i+1)^2$ is Catalan's constant.
We have checked the consistency of our exact finite size data with
these values. 
For instance, the value of $\mu$ can be estimated by considering
the ratio $A_L(\alpha)=P_{L+2}(\alpha)P_L(\alpha)/(P_{L+1}(\alpha))^2$,
which converges to $\mu^2$ at large $L$, and applying to $A_L$ standard
convergence acceleration techniques. In this paper, we found it convenient 
to use a simple linear convergence algorithm which consists, for any
sequence $U_L$ tending at large $L$ to $U_\infty$, in building
new sequences $U_L^{(k)}=\Delta^{(k)}\left[L^k U_L/k!\right]$ for 
increasing integers $k$, where $\Delta^{(k)}$ denotes the $k$-th iteration
of the finite difference operator $\Delta f (L)\equiv f(L+1)-f(L)$.
The sequence $U_L^{(k)}$ is expected to tend faster to $U_\infty$ for larger
$k$ as long as $k$ is kept reasonably small (in practice we used mostly
$k=4$ and went as far as $k=11$ for the jamming probability of Sect.3.2 below).
For instance, if $U_L$ has a Taylor expansion in $1/L$, then the Taylor
expansion of $U_L^{(k)}$ has its first $k$ correction terms vanishing.

\figtex{Estimate for the area entropy $\mu$ at $\alpha=\pm 1$ from the
data of $P_L(\alpha)$ for $L$ up to $60$.}{muestimate2.tex}
\figlabel\muestimate
The estimates for $\mu$ when $\alpha=+1$ 
and $\alpha=-1$ are plotted in Fig.~\muestimate\ as a function of
$L$. We obtain the value $\mu=3.2099123(1)$, fully consistent with
the exact analytic expression above.

More interesting is the value of the exponent $\tilde\gamma$. Here we shall be 
interested only in real values of $y$ in the range $[0,4]$. This corresponds
to taking complex values of $\alpha$ on the unit circle.
The value of $\tilde\gamma$ can be estimated from the quantity
$B_L=-(L^3/2)\ln\left(P_{L+3}(\alpha)(P_{L+1}(\alpha))^3
/(P_{L}(\alpha)(P_{L+2}(\alpha))^3)\right)$, which converges 
to $\tilde\gamma(\alpha)$ at large $L$. Again the results are improved
by use of our convergence acceleration algorithm. 

\figtex{Estimates of the exponent $\tilde\gamma(\alpha)$ for, from
top to bottom, $\alpha=1$, $\alpha=\exp(\pm {\rm i} \pi/4)$,
$\alpha=\pm {\rm i}$, $\alpha=\exp(\pm 3 {\rm i} \pi/4)$ and $\alpha=-1$, 
as obtained from the data of 
$P_L(\alpha)$ for $L$ up to $60$.}{gamestimate2.tex}
\figlabel\gamestimate
Fig.~\gamestimate\ displays 
the estimated values of $\tilde\gamma$ as a function of $L$ for the various 
values of $\alpha$ corresponding to $y=0$ ($\alpha^4=1$) and $y=4$ 
($\alpha^4=-1$). We obtain the asymptotic estimates
\eqn\estione{\eqalign{&\tilde\gamma(1)= 0.749999(1)\cr
&\tilde\gamma(-1)= -0.249999(1)\cr
&\tilde\gamma({\rm i})= \tilde\gamma(-{\rm i})=0.000000(1)\cr}}
for the four roots of $\alpha^4=1$ and 
\eqn\estitwo{\eqalign{&\tilde\gamma(\exp({\rm i}\pi/4))
=\tilde\gamma(\exp(-{\rm i}\pi/4))=
0.3125002(2)\cr &\tilde\gamma(\exp(3{\rm i}\pi/4))=
\tilde\gamma(\exp(-3{\rm i}\pi/4))=-0.187499(1)\cr}}
for the four roots of $\alpha^4=-1$. From these estimates, we conjecture
the exact values 
\eqn\conj{\eqalign{&\tilde\gamma(1)= 3/4\cr &\tilde\gamma(-1)= -1/4 \cr 
& \tilde\gamma({\rm i})= \tilde\gamma(-{\rm i})=0 \cr
&\tilde\gamma(\exp({\rm i}\pi/4))=\tilde\gamma(\exp(-{\rm i}\pi/4))=5/16 
\cr
& \tilde\gamma(\exp(3{\rm i}\pi/4))=\tilde\gamma(\exp(-3{\rm i}\pi/4))=
-3/16 \ .\cr}}

\figtex{Estimated value of $\tilde\gamma(\alpha)$ for $\alpha=\exp({\rm i}
\pi e)$ with $0\le e\le 2$, as obtained from the data of consecutive values of
$P_L(\alpha)$ for $L$ near $60$ (red line). The dashed green line indicates the 
conjectured exact result $\tilde\gamma(\alpha)=3/4-e(2-e)$.}{gamalpha2.tex}
\figlabel\gamalpha
More generally, Fig.~\gamalpha\ displays the estimate 
of $\tilde\gamma(\alpha)$ for all values
of $\alpha$ on the unit circle. These data are fully consistent with
the following analytic expression for $\tilde\gamma$:
\eqn\valtildgamma{\tilde\gamma(\alpha)= {3\over 4}-e(2-e)\quad ,
\quad \alpha=\exp({\rm i} \pi e)\quad ,\quad 0\le e \le 2\ .}
This form for the exponent, quadratic in the phase $e$, 
is frequently encountered in models with statistical weights associated
to loops and will be assumed exact in all subsequent analysis.

{}From Eq.~\factorZ, we immediately deduce from our data the exact generating
function $Z_L(y)\equiv Z(y;i_0)$ for a $(2L+1)\times (2L+1)$ odd white grid with
$i_0$ in the center for $L$ up to $60$. For instance, we get
\eqn\firstZval{\eqalign{
Z_1(y)&= 192+y\cr
Z_2(y)&= 557568000+10474560\,y+y^2\cr
Z_3(y)&= 19872369301840986112+647704492383277056\,y+1642581444224\,y^2+y^3\ .\cr}}
Note that $Z_{60}(0)$ and $Z_{60}(4)$ are $7323$ digit numbers!
{}From Eqs.~\factorZ\ and \scalP, we have the following large $L$ behavior for 
$Z(y;i_0)$
\eqn\scalZ{Z(y;i_0)\sim c(y) { \mu^{(2L+1)^2} \lambda^{4 L} \over 
L^{\gamma (y)}}\ ,} 
where $\gamma(y)=\tilde\gamma(\alpha)+\tilde\gamma(-\alpha)+
\tilde\gamma({\rm i}\alpha) +\tilde\gamma(-{\rm i}\alpha)$ 
with $\alpha$ as in Eq.~\alphatoy\ and
$c(y)=\tilde c(\alpha)\tilde c(-\alpha)\tilde c({\rm i}\alpha)
\tilde c(-{\rm i}\alpha)/\mu$.

\figtex{A direct estimate of the exponents $\gamma(0)$ and $\gamma(4)$ 
obtained from the data of $Z_L(0)$ and $Z_L(4)$ for $L$ up to 
$60$.}{truegamest2.tex}
\figlabel\truegamest
In particular, for $y=0$ ($\alpha^4=1$) and $y=4$ ($\alpha^4=-1$), we obtain
$\gamma(0)=3/4-1/4+0+0=1/2$ and 
$\gamma(4)=5/16+5/16-3/16-3/16=1/4$, namely:
\eqn\scalZZ{\eqalign {Z_{\rm tree} &\sim c(0) 
{ \mu^{(2L+1)^2} \lambda^{4 L}
 \over L^{1/2}} \cr Z_{\rm web}(i_0) &\sim c(4) 
{ \mu^{(2L+1)^2} \lambda^{4 L} \over L^{1/4}}\ . \cr }} 
Note that the first exponent (1/2) can be obtained directly from the
asymptotics of the closed product formula for $Z_{\rm tree}$ \DUDA.
As shown in Fig.~\truegamest, these values are corroborated by a direct 
estimate of $\gamma$ from the exact values of $Z_L(0)$ and $Z_L(4)$ for
$L$ up to $60$. We indeed estimate $\gamma(0)=0.499999(1)$ and 
$\gamma(4)=0.250001(1)$.

Taking the ratio of $Z_{\rm tree}$ and $Z_{\rm web}(i_0)$, 
we deduce that the delocalization probability
${\cal P}_L$ of a vacancy at the center of a $(4L+1)\times (4L+1)$ square
decays at large $L$ as 
\eqn\scalcalP{{\cal P}_L \simeq {1\over L^{1/4}}\ ,}
with some prefactor $c(4)/c(0)$.  

More generally, we immediately get from the expression \valtildgamma\ (with 
$\alpha$ restricted, without loss of generality, to $0\le e \le 1/2$):
\eqn\valgamma{\gamma (y) = u^2 +{1\over 4}\quad ,\quad y=\left(2 
\cos(\pi u)\right)^2 \quad , \quad -{1\over 2} \le u \le {1\over 2}\ ,}
where we have set $u=1/2-2e$.
This expression can be alternatively obtained from a heuristic argument
based on an effective Coulomb gas description of the model. This
argument is presented in Appendix A.

The degree of localization of the vacancy is also measured by the number 
${\cal L}$ of loops of the spanning web that surround it. 
The average of ${\cal L}$ over all spanning webs reads 
\eqn\aveL{\langle {\cal L} \rangle = y {d\, \over d\,y} {\ln Z(y;i_0)}\ .}
At large $L$, we find from the asymptotic behavior \scalZ\ that
\eqn\scalL{\langle {\cal L} \rangle \sim - y {d\, \over d\,y} 
\gamma(y) \  \ln L = {u \over \pi \tan (\pi u)}\ \ln L\ ,}
where we have used the explicit form \valgamma\ of $\gamma(y)$ with 
$y=(2 \cos (\pi u))^2$.
For instance, when $y=4$, (i.e. $u\to 0$), we find that the average 
over all possible dimer configurations of the number of loops around
the vacancy scales as 
\eqn\scaLdimer{\langle {\cal L} \rangle \sim {1\over \pi^2} \ln L}
at large $L$.

\figtex{The exact probabilities for having ${\cal L}$ loops in a spanning web
with $y=4$ for an odd white grid of size $(2L+1)\times (2L+1)$ with
the vacancy in the center. We represent only the cases of ${\cal L}=0,1,2$
as the other probabilities are negligible.}{probloop2.tex}
\figlabel\probloop

From the exact values of $Z_L(y)$, we also have access to the exact
probability for spanning webs to have ${\cal L}$ loops on a finite
square grid with the vacancy in the center.
Fig.~\probloop\ displays these probabilities in the case $y=4$ for 
${\cal L}=0$ (which is nothing but ${\cal P}_L$) and ${\cal L}=1,2$, with $L$ up
to $60$. Note that for this range of grid sizes, the
probability that ${\cal L} \geq 3$ is negligible (less than $10^{-5}$). 
All these probabilities are expected to eventually decay as $L^{-1/4}$
at large $L$.

\subsec{Asymptotic size distribution}

We have established so far that, for a finite grid, the 
delocalization probability ${\cal P}_L$
decreases with the system size $L$ as a power law with exponent $1/4$. 
Let us now see how to extract from this result the value of   
two other exponents which characterize the possible motion of a single vacancy 
in a sea of dimers on an infinite grid. A first exponent, 
$\delta$, characterizes the size 
distribution of sites accessible to 
a vacancy. More precisely, let us consider again a grid of size 
$(4L+1)\times (4L+1)$ covered by dimers with a vacancy at the center 
and consider the probability $p_L(s)$ that the tree component of the 
associated spanning web has $s$ vertices. We expect this probability to tend
at large $L$ to some limiting law $p(s)=\lim_{L\to \infty}p_L(s)$, with
a {\it finite value} for all positive integers $s$. 
This asymptotic distribution
should be universal in the sense that it should not depend on the
precise initial position of the vacancy in the bulk or on the imposed
boundary conditions. 
\figtex{The probability $p_L(1)$ that the vacancy at the center of
a $(4L+1)\times (4L+1)$ grid is strictly jammed.}{probfullyjammed.tex}
\figlabel\fullyjammed

Of particular interest is the value of $p(1)$ which measures the probability
that the vacancy is fully jammed. This value is easily estimated from the 
exact values of $p_L(1)$ at finite $L$. For $s=1$, the tree component of the 
spanning web is reduced to the single vertex $i_0$. In the determinant
formulation of Sect.~2.3, this simply means that no red edge points toward
$i_0$. Imposing $s=1$ therefore amounts to changing $d_i\to d_i-1$ in 
Eq.~\modlap\ for the diagonal terms associated with the four neighbors
of $i_0$ on the grid. Again we can rely on the fourfold symmetry
to reduce by four the size of the matrix involved. We have computed 
$p_L(1)$ for $L$ up to $50$, as plotted in Fig.~\fullyjammed. By applying our
convergence algorithm, we estimate $p(1)=0.10786437626904951198(1)$ from
which we conjecture the exact value
\eqn\estipzero{p(1)={57\over 4}-10 \, \sqrt{2}\ .}
This value was identified thanks to Plouffe's Inverter \Plou\ applied on the
first $10$ digits of $1/\sqrt{p(1)}$. We then verified that it indeed 
reproduces all $20$ digits of $p(1)$ above ! It would be nice 
to have an analytic proof of this result. 

We have also computed
$p_L(2)$ exactly for $L$ up to $34$, from which we estimate 
$p(2)=0.055905353801942(1)$. We have not been able to guess an exact
expression for $p(2)$. 

At large $s$, the distribution $p(s)$ should behave as
\eqn\scalpofs{p(s) \simeq s^{-\delta}}
with an exponent $\delta$ that we shall now compute. 
First, since the delocalization probability ${\cal P}_L$ tends to $0$
at large $L$, it follows that the vacancy in an 
infinite system is always localized. This implies that the
distribution $p(s)$ is normalized to $1$, namely $\sum_{s=1}^{\infty}p(s)=1$. 

To evaluate $\delta$, we consider the probability $1-{\cal P}_L$ that
the vacancy in a finite grid is {\it localized}, i.e. cannot reach the 
boundary of the grid. At large $L$, we can indeed estimate this probability as
the probability that $s$ remains less than a maximal size of order
$4L^2$, namely
\eqn\estiprob{1-{\cal P}_L\sim \sum_{s=1}^{4L^2} p(s) =1 -
\sum_{s=4L^2+1}^{\infty} p(s) \sim 1- \int_{4L^2}^\infty p(s) d\,s \sim 1- 
{\rm const.}\ 
L^{2(1-\delta)}}
Comparing with \scalcalP, one finds
\eqn\deltarel{2(1-\delta)=-{1\over 4}\ , }
from which we deduce
\eqn\deltaval{\delta={9\over 8}\ .}
Despite the fact that the vacancy is localized on a finite tree, the result 
that $\delta <2$ implies that the mean size $\langle s \rangle$ of this tree
nevertheless {\it diverges}. For some observables, the vacancy behaves 
{\it on average} as if it were delocalized. In this sense, our model possesses
a rather unusual property which we may call {\it weak localization}. 

\subsec{Diffusion exponent}

As an illustration of this unusual property, let us now discuss 
the dynamics of diffusion of a single vacancy in a dense sea of dimers.
As we discussed, the motion of the vacancy is induced by a dimer sliding 
axially into it, resulting in a jump of the vacancy by two lattice spacings.
In the spanning web language, this simply amounts to moving the vacancy 
to one of its neighbors on the tree component of the spanning web.
A natural choice of dynamics consists in choosing, at each time step,
one of the four neighbors of the vacancy on the odd white grid at random,
and moving the vacancy to that neighbor if possible, i.e.
if this neighbor belongs to the tree component of the spanning web.
We then define $S(t)$ as the total number of sites visited
by the vacancy after $t$ steps. Although $S(t)$ is clearly bounded by
the size $s$ of the tree component of the spanning web, we expect from
our weak localization property that it can become on average 
arbitrarily large, with a large time behavior of the form
\eqn\aveSt{\langle \bar S(t)\rangle \sim k t^{\eta}\ ,}
with some constant $k$.
Here we first average over all possible motions of the vacancy
up to time $t$ ($\bar S(t)$) for a fixed initial sea of dimers (with the
vacancy in the center) and then average over uniformly chosen realizations 
of this initial dimer configuration. 

It is natural to compare this diffusion to that of a particle diffusing 
with the same dynamical rules on a uniformly chosen infinite spanning tree. 
This latter diffusion 
is characterized by a similar exponent $\eta_0$ with
\eqn\aveSt{\langle \bar S(t)\rangle_0 \sim k_0 t^{\eta_0}\ ,}
where the average $\langle \cdot\rangle_0$ is now taken over all
possible (infinite) spanning trees. 

A simple scaling argument relates the exponents $\eta$ and $\eta_0$. 
Denoting by $s$ the size of the 
tree component of the spanning web for a fixed dimer configuration,
we expect that $\bar S(t) \sim k_0 t^{\eta_0}$ with the infinite spanning
tree diffusion exponent $\eta_0$ as long as $k_0 t^{\eta_0}$ is less
than $s$, while it saturates at $\bar S(t) \sim s$ for longer times.
Using Eq.~\scalpofs\ for the distribution of sizes $s$, we can estimate
\eqn\estisoft{\langle \bar S(t) \rangle \sim \int_0^{k_0 t^{\eta_0}} ds\, s\, 
p(s) + \int_{k_0 t^{\eta_0}}^\infty ds\, k_0 t^{\eta_0}\, p(s) \simeq
t^{\eta_0(2-\delta)}\ .}
Note that it is crucial that $1<\delta<2$ for the two integrals to
scale in the same way, determined by the $t$ dependent limits of the integrals.
In other words, the weak localization property of our model is essential 
for a well defined scaling. 
Taking $\delta=9/8$, we have the relation
\eqn\alphrel{\eta={7\over 8}\ \eta_0\ ,}
which measures the lowering of the diffusion exponent due to
weak localization. 

\newsec{Monte Carlo simulations}

\subsec{Simulations of spanning webs}

We have also found it revealing to perform numerical simulations of our model.
As we shall see, we find very good agreement for all quantities for which we 
gave analytic predictions. Moreover, the simulations allow us to 
estimate both diffusion exponents $\eta$ and $\eta_0$ independently, as well 
as other dynamical exponents.

To begin with, we have simulated spanning webs using a modification
of the Propp-Wilson cycle-popping algorithm for finding spanning trees 
of an arbitrary directed graph \PW. This algorithm is presented in detail in
Appendix B and works as follows: for 
every vertex of the odd white grid other than the vacancy vertex, we pick 
a possible outgoing arrow at random. If the resulting graph contains one 
or more cycles, we choose a cycle and ``pop'' it by picking a
new random outgoing arrow for every vertex on the cycle.
We then keep popping cycles until no cycle is left, resulting in a spanning 
tree. Propp and Wilson were able to show that this process terminates with 
probability one, and that the final result is independent of the order 
in which cycles are popped and produces trees uniformly distributed in the space of spanning trees. To generate spanning webs, we simply modified this algorithm by not popping cycles that encircle the vacancy. 
The proofs in Ref.~\PW\ generalize straightforwardly to show 
that the modified algorithm results in web configurations uniform in the space 
of spanning webs. More precisely, this algorithm produces spanning
webs with oriented loops, hence corresponding to a weight $y=2$ per loop. 
It is then straightforward to correct for this weight so as to get an
arbitrary value of $y$. 

As a preliminary check, we have confirmed that for a number of system sizes 
less than $50$, the  algorithm produces probabilities for the
number of loops (in practice $0$, $1$ or $2$) in good agreement with the 
exact probabilities obtained for $y=2$ by the methods of Sect.~3.1.

We generated $y=2$ spanning webs on odd white grids of 
size $(2L+1) \times (2L+1)$
with the vacancy at the center (these correspond to dimer grids of 
sizes $(4L+1)\times (4L+1)$). Our system sizes were $2L+1=25$, $35$, $51$, 
$75$, $101$, $151$, $201$, $301$, $401$, $501$, $701$, and $1001$.
The number of spanning webs generated was $25000$ for all system sizes
 $2L+1\leq301$, $30000$ for $2L+1=401$, $10000$ for $2L+1=501$, $3000$ 
for $2L+1=701$, and $1500$ for $2L+1=1101$. The results were then transformed 
into probabilities for $y=4$ spanning webs (dimer packings) by weighting 
configurations with ${\cal L}$ loops by an extra factor $2^{\cal L}$.

\figtex{Fraction of delocalized configurations, ${\cal P}_L$,
as a function of the system size.}{WebPoppingPercentDelocalized2.tex}
\figlabel\WebPercentDelocalized

The delocalization probability is measured by the fraction of spanning trees
among spanning webs. This fraction is shown in Fig.~\WebPercentDelocalized\ as
a function of the system size and is well fit by a power law.
Dropping the two smallest system sizes ($2L+1=25$ and $2L+1=35$),
which are still consistent with a power law fit but are too small 
to be safely in the large system limit, we get a delocalization exponent 
$0.246\pm 0.006$, in agreement with the value $1/4$ predicted in Eq.~\scalcalP.

\figtex{Comparison between the exact finite size values of $p_L(1)$, 
together with its conjectured exact asymptotic value, and the numerical
probability of strict jamming.}{WebPoppingJammed.tex}
\figlabel\Jammed

\figtex{Distribution $p(s)$ for the size $s$ of the tree component as measured
from spanning web simulations with free boundary conditions (green filled
circles) on a $401\times 401$ odd white grid (corresponding to a 
$801\times 801$ dimer grid) with the vacancy at the
center, and directly from dimer simulations with
periodic boundary conditions (red open circles) on a $1101\times 1101$
dimer grid. The fit (blue line) is for free boundary conditions and
tree sizes less than $200$.}{TreeSizeDistribution.tex}
\figlabel\WebTreeDist

As for the distribution $p(s)$, we first checked that the probability 
$p_L(1)$ of a fully jammed state is in good agreement with both the exact
value that we obtained for small system sizes ($L\leq 50$) and with
the asymptotic prediction of Eq.~\estipzero\ for $p(1)$ (see Fig.~\Jammed). 

The distribution $p_L(s)$ for $L=200$ is shown in Fig.~\WebTreeDist.
At this stage, it is important to analyze the effect of the boundary
in a finite size grid, which can be described as follows: whenever 
the tree component of the spanning web touches the boundary, it must
span the whole grid, hence has size $(2L+1)^2$. 
For $s\leq L$, the tree component cannot reach the boundary so we expect 
$p_L(s)$ to be an accurate estimate of the asymptotic $p(s)$. For 
$L< s\leq (2L-1)^2$, $p_L(s)$ should become significantly lower than 
$p(s)$ as $s$ increases since configurations that 
contribute to $p_L(s)$ are required
to avoid the boundary, a constraint that does not exist in the
asymptotic limit. Finally, the only possible value with $s>(2L-1)^2$
is $s=(2L+1)^2$ and corresponds to delocalized configurations, with a
value $p_L\left(\left(2L+1\right)^2\right)={\cal P}_L$ that is not 
directly relevant
to estimating $p\left(\left(2L+1\right)^2\right)$. This value 
is not shown in Fig.~\WebTreeDist. By eye, the distribution appears to be 
a power law over almost four decades. Upon fitting the data however, 
a small but significant curvature appears for tree sizes greater than
$10^{2.5}$, as expected from the above argument.
To obtain the exponent, we thus fit only for tree sizes less than 200 
(trees that cannot reach the boundary), which gives a power law exponent
$\delta=1.122\pm 0.008$, in good agreement with the prediction 
$\delta=9/8=1.125$ of Eq.~\deltaval.
It is surprising that the power law fit of the data extends
all the way down to $s=1$. In other words, the data are well approximated
by a (normalized) pure power law distribution $s^{-9/8}/\zeta(9/8)$.
For $s=1$, it gives a value $1/\zeta(9/8)=0.116\dots$, to be compared
with the exact value $0.108\dots$ of Eq.~\estipzero.

As we mentioned above, we expect that the asymptotic distribution $p(s)$
should not depend on the finite size boundary conditions from which
it is determined. In this vein, we have also generated dimer packings 
with {\it periodic boundary conditions}, for grids of odd linear size with 
a single vacancy. Note that the spanning web construction, central to 
this paper, no longer applies as, in particular, we have lost
the global notion of distinct odd and even white grids in this case,
or even the global notion of bicolorability.
Still, these notions are preserved locally and for localized configurations
with small enough $s$, the set of accessible sites again forms a tree 
whose statistics we can analyze along the same lines as for free boundary 
conditions. 

One advantage of periodic boundary conditions is that one can
quickly generate many configurations, directly in the dimer setting,
with the so-called ``pivot
algorithm" [\xref\PivotAlgorithmOne,\xref\PivotAlgorithmTwo]. 
Since there is no longer the constraint for localized configurations 
of avoiding the boundary, periodic boundary conditions have the
further advantage of reaching larger values of $s$ at fixed $L$, 
leading to smaller finite size effects. For a linear size of $1101$
(corresponding to $L=275$ in the spanning web language), 
the resulting tree size distribution is well fit by a power law 
over a larger range 
(tree sizes $1\leq s\leq 10^{4.5}$) than for spanning webs, with an exponent
$\delta=1.121\pm 0.003$, in good agreement with Eq.~\deltaval.

Looking at system sizes ranging from $101$ to $1101$, in
steps of $100$, we find that the fraction of delocalized configurations 
(now defined as configurations in which the vacancy can reach any vertex) 
has a power law exponent of $0.260\pm 0.005$, again in agreement with
Eq.~\scalcalP.

\subsec{Vacancy diffusion}

\figtex{Number of distinct sites visited by a diffusing
vacancy on a $401\times 401$ uniform spanning tree as a function
of time.}{USTSitesVisited2.tex}
\figlabel\USTsitesvisited

\figtex{Squared displacement of a diffusing vacancy on a $401\times 401$
uniform spanning tree as a function of time.}{USTDiffusion2.tex}
\figlabel\USTdiffusion

Let us now come to the prediction \alphrel\
relating the diffusion exponent $\eta$ for the growth with time
of the number of sites visited by a diffusing vacancy to the
corresponding exponent $\eta_0$ for diffusion on an infinite
spanning tree. To our knowledge, none of these exponents is
known exactly and we therefore simulated both diffusion processes.

For diffusion on uniform spanning trees, the simulations were done 
for free boundary conditions on a $401\times 401$ grid, with the diffusing
vacancy initially at the center of the lattice. At each time-step, the 
vacancy chooses one of the four compass directions at random and attempts 
a move in that direction. If there is a tree edge in that direction
the move is carried out and otherwise, the vacancy 
stays at the same position. This slightly unusual random
walk rule is chosen to mimic the standard Monte Carlo
dynamics for the underlying dimers, where each time-step
consists of an attempted dimer move. This means that
a vacancy that has more possible moves should
on average wait less time before moving. 

An ensemble of $1000$ spanning trees was generated and, for each tree,
the vacancy underwent $10^7$ time steps.  
The resulting graphs for the number of sites visited as a
function of time, and the squared displacement as a function
of time, are shown in Figs.~\USTsitesvisited\ and \USTdiffusion.
There is a relatively large curvature at smaller times
($t<100$), so such times are excluded from the fit.
Both graphs show good power law behavior by eye over 
the last five decades of time.
Closer inspection shows that both graphs do have a small
but significant curvature, which we use to estimate the error
bars of the slope (the statistical errors are negligible).

From Fig.~\USTsitesvisited, 
we obtain $\eta_0=0.61\pm 0.02$, while
from Fig.~\USTdiffusion, 
we get $\theta_0=0.62\pm 0.02$, where we have defined the exponent $\theta_0$
through
\eqn\rtwo{\langle \bar{r^2}(t) \rangle_0 \propto t^{\theta_0}\ ,}
where $r(t)$ is the Euclidean distance of the vacancy to the center of
the grid at time $t$ .
These estimates are consistent with the two exponents being
identical.

\figtex{Number of distinct sites visited by a diffusing vacancy
on a $401\times 401$ uniform spanning web as a function
of time.}{WebSitesVisited2.tex}
\figlabel\websitesvisited

\figtex{Squared displacement of a diffusing vacancy on a $401\times 401$ 
uniform spanning web as a function of time.}{WebDiffusion2.tex}
\figlabel\webdiffusion

We then simulated the diffusion of a vacancy on spanning webs. The spanning
webs were generated with the algorithm described in Sect.~4.1. An
ensemble of $10000$ spanning webs was generated, each of the same size 
$401\times 401$ as for the spanning trees above.
For each web, the vacancy moved for the same total time and with the same
dynamics as before.

Figs.~\websitesvisited\ and \webdiffusion\ 
show the number of sites visited and the squared displacement
in the case of spanning webs. 
Again, we fit only for times $t>100$, and use the curvature
to estimate the error bars.
Good power law behaviors are again seen over five
decades.
We obtain
$\eta=0.54\pm0.03$ and $\theta=0.56\pm0.01$ (with $\theta$ defined
as $\theta_0$ but for spanning webs).
Again, these estimates are consistent with the two exponents being
identical.

Finally, we estimate the ratio $\eta/\eta_0=0.89\pm0.06$, in agreement 
with the prediction $7/8=0.875$ of Eq.~\alphrel.

\newsec{Discussion}

In this paper, we have analyzed the possible motion of an isolated
vacancy in an otherwise fully packed dimer model on the square lattice.
We find that the vacancy exhibits weak localization: the size of its
accessible domain scales as a power law with diverging average size. 
Exact finite size enumerations allowed us to identify several exponents
as well as determine the probability for strict jamming. Some of these
results are still awaiting exact and/or rigorous proofs. We also 
gave a universal relation 
between the exponents $\eta$ and $\eta_0$ for the diffusion on spanning webs 
(vacancy diffusion) and that on spanning trees. We have no prediction 
for their individual values but from the numerical data, $\eta_0$ 
is consistent with a simple value 5/8, which would result in a vacancy 
diffusion exponent $\eta$ equal to $35/64$.

The emergence of spanning webs as generalizations of spanning trees
is quite natural here and one might hope that they will appear in
other physical and mathematical problems. 

There are several directions in which the present work can be extended.
Introducing multiple vacancies will lead to vacancy interactions whose
treatment will almost certainly require more elaborate geometrical 
structures. The power of the analysis in this paper relies heavily
on special features of the square lattice. Extension to other lattices
will require new insight. Key among them is the triangular lattice, 
for which we expect vacancies to be localized, with the size 
distribution of accessible sites now decaying exponentially at a rate 
determined by the known entropy mismatch between dimer configurations 
and spanning tree configurations. If we assume that this size is a good
measure of the extent to which a vacancy perturbs its dimer background,
the localization should be related to the exponential convergence
of the ``monomer-monomer" correlation function to a plateau at large 
separation (deconfinement), as observed in Ref.~\FMS.
Note finally that a random lattice version of the problem looks particularly 
promising for obtaining exact results.

\bigskip
\noindent{\bf Acknowledgments:} 
We thank M. Bauer, C. Boutillier, F. David, Ph. Di Francesco, B. Duplantier, 
W. Krauth, J.-M. Luck and X. Xing for enlightening discussions. 
The authors acknowledge 
support from the Geocomp project, ACI Masse de donn\'ees (E.G), from the 
ENRAGE European network, MRTN-CT-2004-5616 (E.G. and J.B) and from the 
Programme d'Action Int\'egr\'ee J. Verne ``Physical applications of random 
graph theory" (E.G.). M.B would like to acknowledge the hospitality 
of the Service de Physique Th\'eorique of Saclay during the period 
in which this work was carried out.

\appendix{A}{Coulomb gas argument}
Let us present here a heuristic derivation of the relation \valgamma\
based on a Coulomb gas formulation of the problem. 
\fig{(a): The paths of the maze made of a spanning web (black) and its dual 
web (red) form a set of loops (green) that completely encode the 
(unrooted) spanning web. (b): The loops are fully-packed, concentric
around the root of the original spanning web and can be oriented so as 
to follow a Manhattan orientation of the underlying grid (as indicated
by arrows).}{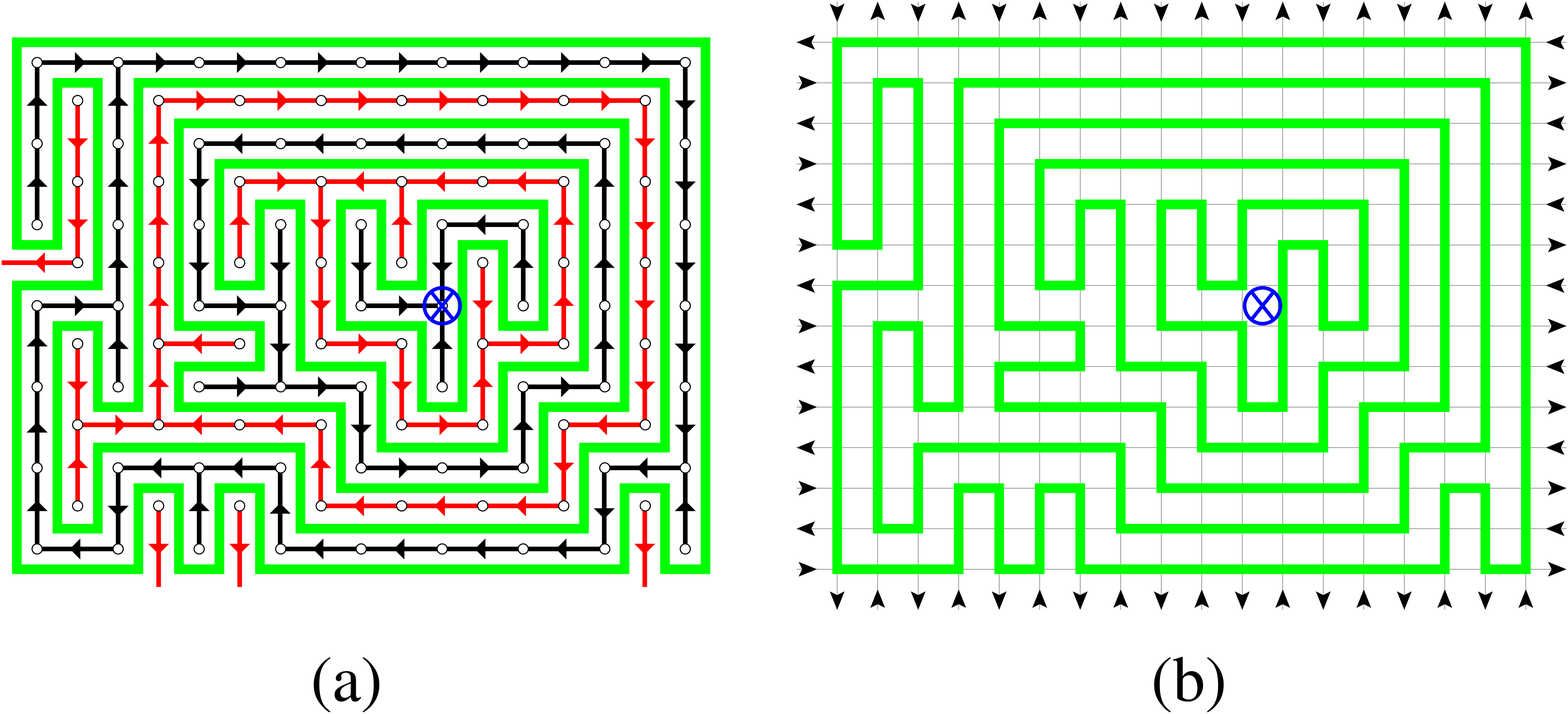}{13.cm}
\figlabel\onloop
First we note that, as illustrated in Fig.~\onloop, any spanning web
configuration with ${\cal L}$ loop components can be alternatively coded 
by a configuration of $2{\cal L}+1$ loops on a square grid dual to the
original dimer grid. These loops are such that: (i) each loop encloses
the root vertex $i_0$, (ii) the loops are self- and mutually-avoiding
and fully-packed, i.e. each vertex of the dual grid is visited by a loop,
(iii) the loops can be oriented consistently so as to follow a Manhattan
orientation on the dual grid. We can relax condition (i) and assign
a weight $n_0$ for each loop that does not enclose $i_0$ 
and a weight $n$ for each loop that does enclose $i_0$. 
The correct statistics with a weight $y$ per loop component of
the spanning web is then recovered by choosing $n_0=0$ and
$n=\sqrt{y}$ (and dividing by the residual weight $\sqrt{y}$
associated with the loop around the tree component). 
The model is expected to lie in the universality class of
the dense O($n_0$) model which can be described by a one-dimensional
height field.\foot{It is known that fully-packed loops are in
general in a different universality class than dense loops. In particular,
on the square lattice, the fully-packed loop universality class is described 
by a $3$-dimensional height variable. For the model at hand with the extra 
constraint (iii) above, however, two of the height components are eliminated, 
leading to an effectively one-dimensional model.}
In terms of this height field, the loop model is mapped onto a Coulomb gas 
model for which various exponents can be obtained exactly \Nien. 
The anomalous weight $n$ for the loops that enclose $i_0$ is properly
accounted for by introducing an electric operator at $i_0$ with charge
$q$ such that $n=2 \cos(\pi (q-u_0))$. 
The dimension of such an electric operator is 
\eqn\dimelec{x={q(q-2 u_0)\over 2 g_0}\ ,}
where $n_0=2\cos (\pi u_0)$ ($0\leq u_0\leq 1$), $g_0=1-u_0$ and
with the determination of $q$ such that $u_0-1/2\leq q\leq u_0+1/2$. 
Here we want $n_0=0$, i.e. $u_0=1/2$ and $n=\sqrt{y}=2\cos (\pi u)$
with $u$ as in Eq.~\valgamma, i.e. $q=u+1/2$. We end up with 
\eqn\valx{x=u^2\,-{1\over 4}\ .}
The dimension $x$ of the electric operator measures in particular 
the algebraic decay as $L^{-x}$ of its average in a finite geometry
of linear size $L$. For a square grid geometry, this average is 
nothing but the ratio $Z(y;i_0)/Z(0;i_0)$ where the denominator is
identified as the partition function of the system in the absence of 
electric operator. We immediately deduce 
the relation $x=\gamma(y)-\gamma(0)$, i.e.
\eqn\newvalgamma{\gamma(y)=u^2 -{1\over 4}+\gamma(0)\ .}
To end the argument and recover Eq.~\valgamma, we simply rely on 
a direct calculation of  $\gamma(0)=1/2$, as obtained in Ref.~\DUDA\ from 
the exact product formula for $Z(0;i_0)=Z_{\rm tree}$.

\appendix{B}{Generating spanning webs via cycle-popping}

In this appendix we explain how to generate a random spanning web
using a ``cycle-popping'' algorithm inspired from the Propp-Wilson
algorithm for the generation of a random spanning tree. More precisely
we extend to random webs the {\tt RandomTreeWithRoot()} procedure
explained in Sect.~6 of Ref.~\PW.
\fig{An example of a configuration (a) obtained by drawing at random an edge
from each vertex but the root $i_0$. A possible configuration (b) obtained
from (a) by popping the magenta cycle. The tree component flowing to $i_0$ 
(here in green) can only grow in the process. To generate spanning webs instead
of spanning trees, we make all cycles winding around $i_0$ (such as the brown
cycle) unpoppable.}{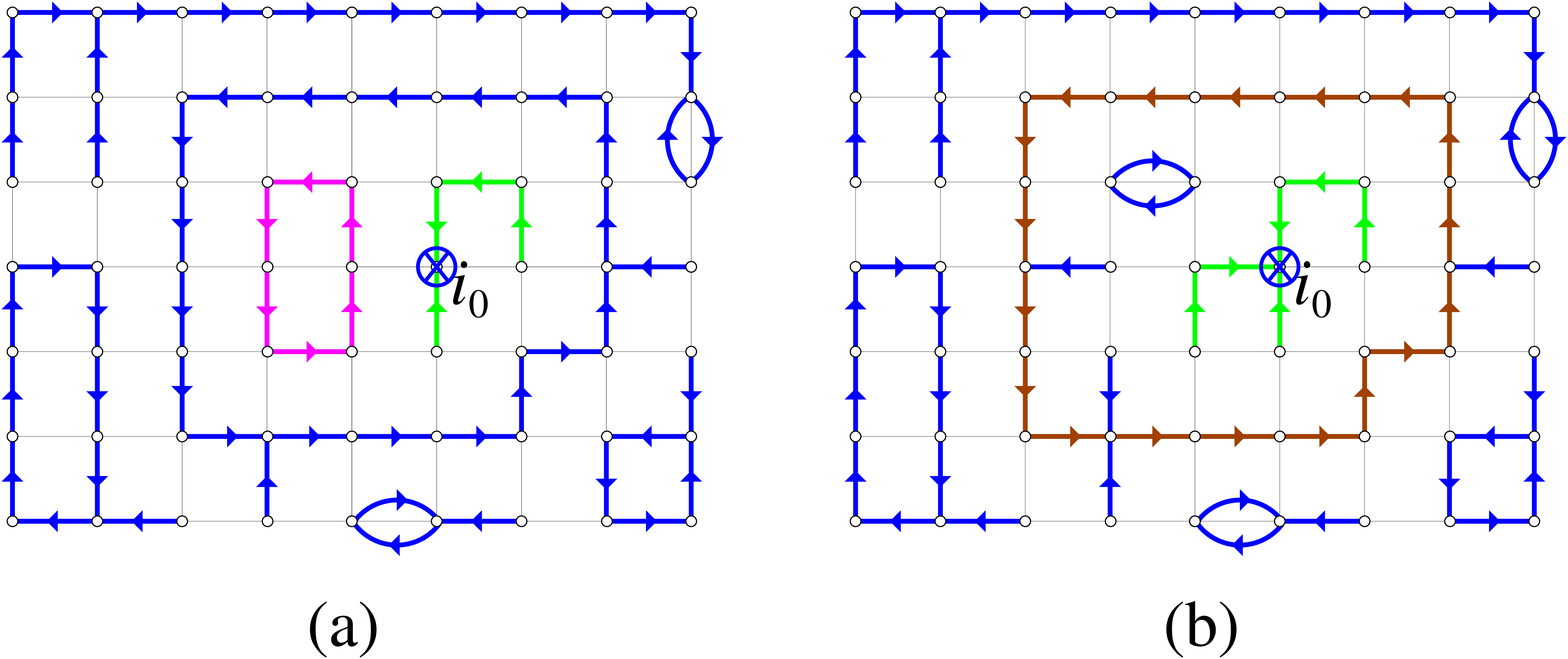}{13.cm}
\figlabel\cyclepop
Let us begin by recalling how the Propp-Wilson algorithm works in our
specific setting. Start with a rectangular grid on which the spanning
tree is to be constructed, with the root $i_0$ at a given position.
At each vertex distinct from $i_0$, draw a random incident edge
uniformly and independently from the other vertices. Pictorially the
edge is marked and oriented away from the vertex : a possible outcome
of this procedure is illustrated in Fig.~\cyclepop-(a). Note the
similarity with the proof of the determinant formula \determinant. By
chance, the graph made of the selected edges can be a spanning tree
rooted at $i_0$ (with edges oriented towards the root), but more
likely it will contain one or more (disjoint) cycles (loops) as in
Fig.~\cyclepop-(a). Cycle-popping consists in choosing an arbitrary
cycle and {\it popping} it, i.e. for each vertex on the cycle, drawing
a new random outgoing edge uniformly and independently of all previous
draws. For instance, Fig.~\cyclepop-(b) shows a possible outcome after
the popping of the magenta cycle in Fig.~\cyclepop-(a). 
The resulting graph can again contain cycles (some preexisting 
and some created by popping): in that case repeat the
procedure by choosing another cycle and popping it. Otherwise the
resulting graph is a spanning tree and the procedure terminates. In
Fig.~\cyclepop, the green edges ``flow'' to the root, hence they
cannot be in a cycle and they must belong to the final tree. 
One then sees that the green graph can only grow at each step,
eventually covering the whole grid.

There are several ways to choose which cycle to pop at each step of
the procedure (one could imagine doing it deterministically or randomly),
but Propp and Wilson have shown that the actual procedure
is (in some precise sense) irrelevant. Furthermore: (i) the
procedure terminates almost surely, i.e. only a finite number of cycles
have to be popped before none remains, and (ii) the resulting spanning tree is
a {\it perfect} sample of the uniform measure on the set of spanning
trees rooted at $i_0$. Propp and Wilson actually deal with a slightly
more general case, not needed here, and compute bounds on the running
time, which establish that this is a very efficient algorithm.

The modification of the algorithm to generate spanning webs is
easy. Perform the procedure as before, except that 
cycles winding around the root $i_0$ are now considered {\it
unpoppable} and stay in the resulting graph. For instance in
Fig.~\cyclepop-(b) the brown cycle would be unpoppable if we were to
generate a spanning web. Only the cycles not winding around the root
are popped, until none remains. Propp and Wilson's analysis (in
Sect.~7 of Ref.~\PW) extends straightforwardly to the case of the
unpoppable cycles, and we find that : (i) the procedure terminates
almost surely, and (ii) the resulting graph is a now a {\it perfect}
sample of the uniform measure on graphs (made out of oriented edges,
with exactly one edge going out of every vertex, except for the root
that has none) whose cycles are all unpoppable. 
One sees easily that such a graph is nothing but a spanning
web rooted at $i_0$, 
as defined in Sect.~2.2, with additional orientations such that 
each edge not belonging to a
cycle ``flows'' towards the root or an attractor cycle. Since each
cycle has two possible orientations that are equally likely, we
conclude that the probability to obtain a given unoriented spanning
web containing $\cal L$ loops is $2^{\cal L}/Z(2,i_0)$, where
$Z(2,i_0)$ is the generating function for spanning webs on the grid at
hand, rooted at $i_0$ and counted with a weight $y=2$ per loop, as
defined in Eq.~\moddet. This is to be contrasted with the weight $y=4$
per loop that arises from the correspondence with dimer configurations
with a vacancy, drawn with uniform probability. Knowing the exact bias, it
is however straightforward to translate the statistical properties of 
spanning webs as measured through this algorithm into dimer statistics.
Moreover, one can correct the bias directly in the algorithm, at the
price of introducing suitable weights for edges along a seam.

In conclusion, we have provided an algorithm for the generation of
random spanning webs belonging to the class of exact or {\it perfect}
algorithms [\xref\PW,\xref\Dbw]. By the
correspondence of Sect.~2.2, it can be used to simulate dimer
configurations on a square lattice with a vacancy. We have not
performed a detailed analysis of its efficiency, but we believe it is
comparable to the original Propp-Wilson algorithm for rooted spanning
trees, and much better than generic CFTP-type algorithms. 
The comparison with
other Monte Carlo algorithms, such as the pivot algorithm
[\xref\PivotAlgorithmOne,\xref\PivotAlgorithmTwo] for the case of
periodic boundary conditions is not so clear:
typically, these are designed with efficiency in mind, but without an 
exact knowledge of their ``randomness''. Both approaches provide 
useful and mutually consistent results in our study.

\listrefs
\end